\documentclass[11pt]{article}
\usepackage{a4wide,amsmath,epsfig}
\usepackage{longtable}
\allowdisplaybreaks[4]
\newcommand{\eff}{\text{eff}}
\begin{document}

\vspace*{0.2cm}

\rightline{\today}
\begin{center}
{\Large \bf On different lagrangian formalisms for vector resonances}\\
{\Large \bf within chiral perturbation theory}\\[1.5 cm]

{\bf Karol Kampf}\footnote{for emails use: {\it surname\/} at ipnp.troja.mff.cuni.cz},
{\bf Ji\v{r}\'{\i} Novotn\'y}\footnotemark[\value{footnote}] and {\bf Jaroslav Trnka}\footnotemark[\value{footnote}]\\[1 cm]
{\it Institute of Particle and Nuclear Physics, Faculty of Mathematics and Physics,}\\
{\it Charles University, V Hole\v{s}ovi\v{c}k\'ach 2, CZ-180 00 Prague 8, Czech
Republic}
\\[0.5 cm]

\end{center}

\vspace*{2.0cm}

\begin{abstract}
We study the relation of vector Proca field formalism and
antisymmetric tensor field formalism for spin-one resonances in
the context of the large $N_C$ inspired chiral resonance
Lagrangian systematically up to the order $O(p^6)$ and give a
transparent prescription for the transition from vector to
antisymmetric tensor Lagrangian  and vice versa. We also discuss
the possibility to describe the spin-one resonances using an
alternative ``mixed'' first order formalism, which includes both
types of fields simultaneously, and compare this one with the
former two. We also briefly comment on the compatibility of the
above lagrangian formalisms with the high-energy constraints for
concrete $VVP$ correlator.
\end{abstract}

\newpage

\setcounter{footnote}{0}

\section{Introduction}

The low energy effective theory of QCD, known as chiral perturbation theory
($\chi PT$)\cite{Weinberg:1978kz,Gasser:1983yg,Gasser:1984gg}, has
made a considerable progress recently by means of the extension of
the calculational scheme to the order $O(p^6)$ (a comprehensive
review of the recent calculations and the most important
achievements can be found in \cite {Bijnens:2006zp} together with a
complete list of references). The successful solution of the
technical problems and increasing number of physical quantities
computed at the next-to-next-to-leading order within the
three-flavour expansion raised an issue of much more detailed
knowledge of the free parameters (known as low-energy couplings
(LECs)) of the $\chi PT$ \thinspace Lagrangian ${\cal L}_\chi $,
\begin{equation}
{\cal L}_\chi ={\cal L}_\chi ^{(2)}+{\cal L}_\chi ^{(4)}+{\cal
L}_\chi ^{(6)},  \label{L_chi}
\end{equation}
in order to make contact of the higher precision $O(p^6)$
calculations with phenomenology. The LECs, representing
nonperturbative parameters of the underlying QCD dynamics
associated with chiral symmetry breaking, are not
uniquely determined from symmetry considerations only, and, at the order
$O(p^6)$, only few  combinations of them are directly accessible
from experimental data. The usual way of thinking about the actual
values of LECs is to associate them with the physics of the
low-lying resonances at the scale $\Lambda _H\sim 1GeV$ via chiral
sum rules (for $O(p^4)$ LECs listed {{\em {e.g.}} in
\cite{Gasser:1983yg}). These relate the LECs to the low-energy
expansion of the QCD correlators at zero momentum and exploit the
information about the high-energy fall off of the correlators
known from the operator product expansion (OPE). The corresponding
spectral functions in the intermediate resonance region are either
taken from experiment (see {{\em {e.g.}}
\cite{Donoghue:1993xb,Davier:1998dz}) or modelled using plausible
theoretical assumptions (see {{\em {e.g.}} \cite{Moussallam95,
knecht}). The considerations based on the large-$N_C$ expansion of QCD
are particularly useful in such a type of reasoning \cite{'tHooft:1973jz}.
In the limit $N_C\rightarrow \infty $, the
correlators of quark bilinears can be calculated in terms of tree
graphs within an effective theory containing infinite tower of
weakly coupled
narrow meson resonances with appropriate quantum numbers. The Lagrangian
${\cal L}_\infty $ of this effective theory (called chiral
resonance Lagrangian) is not known from the first principles,
however, it can be basically constructed on the symmetry grounds
and its free parameters can be related to the phenomenology of the
resonance sector. Up to the order $O(p^6) $ it has the general
form
\begin{eqnarray}
{\cal L}_\infty  &=&{\cal L}_{GB}+{\cal L}_{res}  \nonumber \\
&=&{\cal L}_{GB}^{(2)}+{\cal L}_{GB}^{(4)}+{\cal L}_{GB}^{(6)}+{\cal L}%
_{res}^{(4)}+{\cal L}_{res}^{(6)}\,,  \label{L_infinity}
\end{eqnarray}
where ${\cal L}_{GB}=$ ${\cal L}_{GB}^{(2)}+{\cal L}_{GB}^{(4)}+{\cal L}%
_{GB}^{(6)}$ contains only the (pseudo)Goldstone bosons and ${\cal L}%
_{GB}^{(2n)}$ has the same form as the $O(p^{2n})$ $\chi PT$
\thinspace Lagrangian ${\cal L}_\chi ^{(2n)}$(cf. (\ref{L_chi});
the corresponding LECs are, however, different\footnote{%
Actually, in concrete resonance saturation calculation, these LECs
are treated as negligible at the resonance scale, because, within
the truncated chiral resonance Lagrangian,  they correspond to the
physics at the energy scales higher then the low-lying resonance
states, which are treated as
active degrees of freedom here.}). ${\cal L}_{res}^{(4)}$ and ${\cal L}%
_{res}^{(6)}$ are the resonance Lagrangians of the chiral order
$O(p^4)$ and
$O(p^6)$ respectively\footnote{%
The chiral order of the resonance fields depend on the formalism
used and will be clarified in what follows.}. Integrating out the
resonance fields from the chiral resonance Lagrangian and
expanding to the given chiral order yields an effective $\chi PT$
\thinspace Lagrangian ${\cal L}_\chi $ identified with
(\ref{L_chi}) with LECs expressed in terms of the resonance
parameters and the LECs from ${\cal L}_{GB}$. Schematically, up to
the order $O(p^6)$
\begin{eqnarray}
{\cal L}_\chi  &=&{\cal L}_{GB}+{\cal L}_{\chi ,\,res} \nonumber\\
&=&{\cal L}_{GB}^{(2)}+{\cal L}_{GB}^{(4)}+{\cal L}_{GB}^{(6)}+{\cal L}%
_{\chi ,\,res}^{(4)}+{\cal L}_{\chi ,\,res}^{(6)}\,.
\end{eqnarray}
Here ${\cal L}_{\chi ,\,res}^{(2n)}$ has the same form as ${\cal L}%
_{GB}^{(2n)}$ with LECs depending on the resonance masses and couplings of
${\cal L}_{res}={\cal L}_{res}^{(4)}+{\cal L}_{res}^{(6)}$. Though
in principle ${\cal L}_{res}$ should contain infinite tower of
resonance fields, it usually suffices to include only a finite
number of them (corresponding to the lowest hadronic states in
each channel) to saturate the LECs successfully, as it was shown
in the seminal paper \cite{ecker1} for the $O(p^4)$ LECs. Since
then this idea has been often used in particular cases in order to
estimate also the contribution of the $O(p^6)$ LECs to various
quantities calculated within the $O(p^6)$ $\chi PT$. Quite
recently, the first steps towards a systematic and consistent estimate of the
$O(p^6)$ LECs via resonance saturation have been made in
\cite{CEEKPP, moussallam}, where the general discussion of the
method as well as the comparison of the results with experimental
constraints from the observables of $\pi K$ scattering can be
found. }

However, the Lagrangian ${\cal L}_\infty $, given by
(\ref{L_infinity}), is not unique. The reason is that a general
Lagrangian for spin-1 resonances can be formulated using various
types of fields, the most common ones in this context are the
vector
Proca fields and antisymmetric tensor fields ({\em i.e.} the label
$res\equiv V$ or $T$ in the above formulae), which both transform
homogeneously under a nonlinear realization of the chiral symmetry\footnote{%
In this paper we will not discuss other approaches to the
resonance Lagrangians, namely the ``Massive Yang-Mills''
\cite{Kaymakcalan:1983qq, Gomm:1984at, Meissner:1987ge}, ``Hidden
local symmetry'' \cite {Bando:1984ej,Bando:1987br} and
``Generalized hidden local symmetry'' \cite
{Bando:1985rf,Bando:1987ym,Kaiser:1990yf}. The relation of these
resonance models to the vector and antisymmetric tensor
formulation at $O(p^4)$ is discussed {\em {e.g.}} in Refs.
\cite{ecker2,pallante,Borasoy:1995ds,
Harada:2003jx,Birse:1995yw,Birse:1996hd}}. The equivalence of these
two formulations of the resonance Lagrangian was studied in the
past (cf. references \cite{ecker1,
ecker2,abada,kalafatis,pallante, bijnens, tanabashi}).

As pointed out in \cite{pallante}, there are several levels of
equivalence of the various resonance Lagrangians. The first
corresponds to the level of the chiral resonance Lagrangians
${\cal L}_\infty $. {\em Complete equivalence} here means equality
of the physical observables (Green functions), calculated in the
full chiral resonance theory, after the various coupling constants
in one model are expressed in terms of the coupling constants of
the other model. In other words, there exists a unique transformation
connecting the coupling constants in both chiral resonance
Lagrangians, which relates the results of both calculations.
A weaker form of such an equivalence, the {\em incomplete} one,
ensures equality of the observables modulo terms of the higher
chiral order ($O(p^8)$ in our case).

The second level compares the Lagrangians ${\cal L}_{res}$ only.
It means equivalence of the contributions to the corresponding
effective chiral
Lagrangians ${\cal L}_{\chi ,\,res}={\cal L}_{\chi ,\,res}^{(4)}+{\cal L}%
_{\chi ,\,res}^{(6)}$ obtained after integrating out the resonance
fields. Again on this level one can distinguish two cases \cite{pallante}.
{\em Strong
equivalence} is defined as the equality of contributions to the
${\cal L}_{\chi ,\,res}$ in the sense that both formulations yield the same
contributions to the effective chiral couplings when expressing the various
parameters in one resonance Lagrangian ${\cal L}_{res}$
in terms of the parameters of the other one. Strong
equivalence ensures therefore equality of the chiral expansion of
the contribution from ${\cal L}_{res}$ to physical observables up
to the order $O(p^6)$. {\em Weak equivalence} means equality of
the effective $O(p^6)$ chiral Lagrangians ${\cal L}_{\chi ,\,res}$
only when additional contact terms independent of the resonance
fields of the form ${\cal L}_{GB}^{(4)}+{\cal L}_{GB}^{(6)}$ (with
particularly adjusted LECs) are appended to one of the resonance
Lagrangians.

The {\em complete} equivalence of ${\cal L}_\infty $ expressed in
terms of
vector and antisymmetric tensor fields, was first proved up to the order
$O(p^4)$ in Ref. \cite{ecker2} using the high-energy constraints to
find the relation between the couplings in both approaches. More
precisely, the general Lagrangian with antisymmetric tensor fields
\begin{equation}
{\cal L}_{T,\infty }={\cal L}_{GB}^{(2)}+{\cal L}_{\,T}^{(4)}
\end{equation}
was shown to be completely equivalent to vector field Lagrangian
\begin{equation}
{\cal L}_{V,\infty }={\cal L}_{GB}^{(2)}+{\cal L}_{GB}^{(4)}+{\cal L}%
_{\,V}^{(6)}
\end{equation}
with particular choice of ${\cal L}_{GB}^{(4)}$ and ${\cal
L}_{\,V}^{(6)}$. Within the above classification, however, ${\cal
L}_{\,T}^{(4)}$ and this particular ${\cal L}_{\,V}^{(6)}$ are
only {\em weakly} equivalent, because the corresponding ${\cal
L}_{\chi ,\,T}^{(4)}$ and ${\cal L}_{\chi ,\,V}^{(4)}$ differs by
contact terms from ${\cal L}_{GB}^{(4)}$. These contact terms
were found to be necessary to correct the possible wrong
high-energy behaviour of certain observables calculated within the
vector formalism. The role of such contact terms ensuring
the positivity of the Hamiltonian was demonstrated in
\cite{abada, kalafatis}, where also the proof of the above
equivalence in the hamiltonian formalism was given. Later on the
equivalence was proved using path integral methods in Ref.~\cite{bijnens}
(cf. also \cite{Borasoy:1995ds} for critical remarks)
and in Ref. \cite{tanabashi} (cf. also \cite{Harada:2003jx}).

With restriction to the anomalous sector of ${\cal L}_{\chi,\,res}$, the
{\em weak} equivalence of ${\cal L}_{\,T}^{(4)}+{\cal L}_{\,T}^{(6)\,{\rm odd%
}}$ and ${\cal L}_{\,V}^{(6)}+{\cal L}_{\,V}^{(6){\rm odd}}$ was
established \cite{pallante} (where ${\cal L}_{\,T}^{(4)}$ and
${\cal L}_{\,V}^{(6)}$ were the same as in the previous case and
${\cal L}_{\,T}^{(6)\,{\rm odd}}$ and ${\cal L}_{\,V}^{(6){\rm
odd}}$ were constructed according to the ``minimal coupling''
hypothesis), however, the complete equivalence was not studied.

In this paper, we try to enlarge the analysis of the equivalence
of vector Proca field formalism and antisymmetric tensor formalism
systematically up to the order $O(p^6)$ and give a transparent
prescription for the transition from vector to antisymmetric
tensor Lagrangian and vice versa. We also study the possibility
to describe the spin-one resonances using an alternative ``mixed''
first order formalism formulated in terms of both
types of fields simultaneously and compare this one with the former two.
We also briefly comment on the compatibility of
the above lagrangian formalisms with the high-energy constraints
for concrete QCD correlator.

The paper is organized as follows. In Section \ref{S2} we first fix our
notation and briefly review the Proca field and antisymmetric
tensor field formalisms. Section \ref{S3} is devoted to the detailed
analysis of the equivalence of both approaches. As an explicit
example of the various levels of equivalence and as an
illustration of the transition prescription from antisymmetric
tensor to vector form we use the $VVP$ correlator in Section \ref{S4}.
Alternative first order description of spin-one resonances is
presented in Section \ref{FO}, where also the relation to the former two
approaches is clarified and illustrated using $VVP$ correlator.
Brief summary is given is Section \ref{S6}. Formal properties of the
first order formalism, as well as some technical details are
included in the Appendix~\ref{appprop}.

\section{Basic concepts and notation}\label{S2}

In what follows, we restrict our discussion only to one multiplet of the vector
resonances, for definiteness we assume $1^{--}$ nonet. In this section we
briefly review two standard possible representation of this particle states
using Proca vector fields, which we denote generically $V$ $\equiv V_\mu ^a$
and the antisymmetric tensor fields, for which we use the general symbol
$R\equiv R_{\mu \nu }^a$. In what follows, in order to avoid cumbersome
expressions, we shall systematically suppress both the Lorentz and group
indices and use a condensed ``dotted'' and ``bracket'' notation for their
respective contractions. Within this notation, we write for two
generic four-vectors $A_\mu ^a$ and $B_\nu ^b$%
\[
(A\cdot B)\equiv A_\mu ^aB^{a\mu }
\]
and, more generally, whenever a dot appears between two tensors, it means
contraction over the adjacent Lorentz indices. Also whenever two or more
objects with group indices appear in a given sequence, we tacitly assume
contraction of the adjacent group indices. Within this convention, {\em e.g.}
\[
(V\cdot K\cdot V)\equiv V_\mu ^aK^{ab\mu \nu }V_\nu ^b\,.
\]
For generic tensors we employ ``$:$'' for a pair of contracted
antisymmetric indices, {\em i.e.}
\begin{eqnarray*}
R:J &\equiv &R_{\mu \nu }J^{\mu \nu } \\
R:J::RR &\equiv &R^{\mu \nu }J_{\mu \nu \,\,\alpha \beta \,\,\kappa \lambda
}R^{\alpha \beta }R^{\kappa \lambda }\,.
\end{eqnarray*}
We also mark by hat symbol an antisymmetrization, {\em e.g}. for a generic
Lorentz rank two tensor $C_{\mu \nu }^{a\ldots }$ we write
\[
\widehat{C}\equiv C_{\mu \nu }^{a\ldots }-C_{\nu \mu }^{a\ldots }
\]
and generally for antisymmetrization of two adjacent Lorentz indices
\[
\widehat{AB}\equiv A_{\ldots \mu }B_{\nu \ldots }-A_{\ldots \nu }B_{\mu
\ldots }\,.
\]
As an example of these rules, we can abbreviate $D_\mu ^{ab}V_\nu ^b-D_\nu
^{ab}V_\mu ^b$, where $V$ is the Proca field and $D\equiv D_\mu ^{ab}$ is
the usual covariant derivative as, $\widehat{DV}$.

\subsection{Proca field formalism}

Let us take into account just the interaction terms linear and quadratic in
the resonance fields and write the Lagrangian for the vector-field
representation in the following symbolic form
\begin{equation}
{\cal L}_V=-\frac 14(\widehat{V}:\widehat{V})+\frac 12m^2(V\cdot
V)+(J_1\cdot V)+(J_2:\widehat{V})+\frac 12(V\cdot K\cdot V)+(V\cdot J_3:%
\widehat{V})\,.  \label{L_V}
\end{equation}
With a slight abuse of the above notation we have further introduced here
$\widehat{V}=\widehat{DV}$ . Here and in what follows, we will often use for
a generic external source built of chiral blocks a symbol $J_i$, where $i$ indicates a number of
Lorentz indices.

The most general form of the Lagrangian (\ref{L_V}) to lowest
non-trivial order in the chiral expansion was first constructed in \cite
{Prades94} and often used in the literature to estimate the $O(p^6)$
low-energy constants (see {\em e.g.} in \cite{knecht}). Explicit expressions for
the external sources $J_i$ , $i=1,2,3$ and $K$ in terms of the usual basic
chiral building blocs are given in the Appendix \ref{apJ}. Here we only mention the
chiral order of the sources
\begin{eqnarray}
J_1 &=&O(p^3)\,, \nonumber \\
J_2 &=&O(p^2)\,, \nonumber \\
J_3 &=&O(p)\,, \nonumber \\
K &=&O(p^2)\,.
\end{eqnarray}

Let us make several notes here. First, the interaction terms linear in the
resonance fields start at $O(p^3)$. This suggests to assign to the field
$V_\mu $ the chiral order $O(p^3)$, provided we are interested only in the
resonance contributions to the low energy constant of the $O(p^6)$ chiral
Lagrangian. In this case we have the decomposition
\begin{equation}
{\cal L}_V={\cal L}_V^{(6)}+{\cal L}_V^{(8)}\,,
\end{equation}
where
\begin{eqnarray*}
{\cal L}_V^{(6)} &=&\frac 12m^2(V\cdot V)+(J_1\cdot V)+(J_2:\widehat{V})\,, \\
{\cal L}_V^{(8)} &=&-\frac 14(\widehat{V}:\widehat{V})+\frac 12(V\cdot
K\cdot V)+(V\cdot J_3:\widehat{V})
\end{eqnarray*}
and the corresponding effective $O(p^6)$ chiral Lagrangian can be then obtained
by integrating out the resonance field to the desired chiral order. Because
the integration is gaussian, this effectively means inserting
the solution of the equation of motion to the lowest non-trivial order $O(p^3)$
\begin{equation}
V^{(3)}=-\frac 1{m^2}\left( J_1-2D\cdot J_2\right)
\end{equation}
in ${\cal L}_V^{(6)}$. The result is
\begin{eqnarray}
{\cal L}_{\chi ,V}^{(6)} &=&-\frac 1{2m^2}\left( \left( J_1-2D\cdot
J_2\right) \cdot \left( J_1-2D\cdot J_2\right) \right)  \nonumber \\
&=&-\frac 1{2m^2}\left( J_1\cdot J_1\right) +\frac 2{m^2}\left( D\cdot
J_2\cdot J_1\right) +\frac 2{m^2}\left( D\cdot J_2\cdot J_2\cdot
\overleftarrow{D}\right)  \label{L6V}
\end{eqnarray}
As pointed out in \cite{ecker2}, the contributions to the $O(p^4)$
LEC are not generated, unless extra contact terms are added to the
Lagrangian. On the other hand, the interaction terms contained in the ${\cal %
L}_V^{(8)}$ give contributions only to the $O(p^8)$ chiral Lagrangian and
could be therefore ignored. Note also that, in principle, higher
derivative terms as well as terms cubic in the resonance fields can be added
to the Lagrangian, however (counting $V=O(p^3)$), these are of higher chiral order
(at least $O(p^8)$) and are therefore irrelevant for our consideration.
Nevertheless, both types of the additional terms mentioned above might be
necessary in order to ensure the proper short-distance behaviour of certain
Green functions dictated by OPE, cf. {\em e.g.} \cite{ecker2}, \cite{knecht}.

Last note concerns the form of ${\cal L}_V^{(6)}$. Using integration by
parts, which leads to the {\em completely} equivalent Lagrangian on the
resonance level, one can of course get rid of the term $(J_2:\widehat{V})$
redefining\footnote{%
Provided the source $J_k$ is expressed in terms of the complete operator
basis
\[
J_k=\sum_i\lambda _k^{(i)}{\cal O}_{(i)},
\]
this modification and similar ones in what follows lead only to the
redefinition of the corresponding couplings $\lambda _k^{(i)}$.}
$J_1\rightarrow J_1-2D\cdot J_2$ (cf. also (\ref{L6V})). However, keeping $J_2$
makes the comparison with antisymmetric tensor formulation (and
identification of the contact terms) more transparent.

\subsection{Antisymmetric tensor field formalism}

The Lagrangian in the antisymmetric tensor formulation has the following form
\begin{align}\label{L_T}
{\cal L}_T = &-\frac 12(W\cdot W)+\frac 14m^2(R:R) \\
&+(J_1\cdot W)+(J_2:R)+(W\cdot J_3:R)+(R:J_4:R)+(R:J_5\cdot D:R)+(R:J_6::RR)\,,
\notag
\end{align}
where $R^{a\alpha \mu}$ is the antisymmetric tensor field and
\begin{equation}
(W)^{a\mu }=(D\cdot R)^{a\mu }=D_\alpha ^{ab}R^{b\alpha \mu }\,.  \label{W}
\end{equation}
The chiral order of the sources are
\begin{eqnarray}
J_1 &=&O(p^3)\,, \nonumber\\
J_2 &=&J_2^{(2)}+J_2^{(4)}=O(p^2)+O(p^4)\,, \nonumber\\
J_3 &=&O(p)\,, \nonumber\\
J_4 &=&O(p^2)\,, \nonumber\\
J_5 &=&O(p)\,, \nonumber\\
J_6 &=&O(p^0)\,.
\end{eqnarray}

The most general form of the external sources $J_i$, $i=1,...,6$ with even
intrinsic parity, (which are generally different from the Proca field
formalism) has been constructed in \cite{CEEKPP} (for the complete list of
$SU(3)$ breaking terms see also \cite{moussallam}). The sources with odd
intrinsic parity involving vertices with two resonances and one pseudoscalar
and with one resonance, one pseudoscalar and one vector external source can be
found in \cite{RuizFemenia, pallante}.

Again, as in the case of the vector-field representation, one can use
integration by parts to reduce the form of the above Lagrangian and
eliminate the sources $J_1$ and $J_3$; this leads to the redefinition of the
other sources\footnote{%
As in the vector field formulation we get in this way a {\em completely}
equivalent Lagrangian.}, namely, in the symbolic notation (here $g\equiv
g_{\mu \nu }$ is the metric tensor)
\begin{eqnarray}
J_2 &\rightarrow &J_2-\frac 12\widehat{DJ}_1\,,  \nonumber \\
J_4 &\rightarrow &J_4-\frac 12\widehat{DJ}_3\,,  \nonumber \\
J_5 &\rightarrow &J_5-\frac 12\widehat{gJ}_3\,.  \label{byparts}
\end{eqnarray}
In \cite{CEEKPP}, the only nonzero sources taken into account are $J_2$,
$J_4 $ and $J_6$, in \cite{RuizFemenia} also terms with $J_1$ and $J_5$ are
considered.

Because now the linear source $J_2$ starts at $O(p^2)$, the resonance field
$R$ can be counted as $O(p^2)$ and the full Lagrangian splits according to
the chiral order to
\begin{equation}
{\cal L}_T={\cal L}_T^{(4)}+{\cal L}_T^{(6)}\,,
\end{equation}
where
\begin{eqnarray*}
{\cal L}_T^{(4)} &=&\frac 14m^2\left( R:R\right) +\left( J_2^{(2)}:R\right)\,,
\\
{\cal L}_T^{(6)} &=&-\frac 12\left( W\cdot W\right) +\left(
J_2^{(4)}:R\right) +\left( J_1\cdot W\right) +(W\cdot J_3:R)+(R:J_4:R) \\
&&+(R:J_5\cdot D:R)+(R:J_6::RR)\,.
\end{eqnarray*}
Integrating out the resonance field to $O(p^6)$ needs to insert the solution
of the $O(p^4)$ equation of motion
\begin{equation}
R^{(2)}=-\frac 2{m^2}J_2^{(2)}
\end{equation}
in ${\cal L}_T^{(4)}$ and ${\cal L}_T^{(6)}$. The effective $O(p^4)$ and
$O(p^6)$ chiral Lagrangian then reads
\begin{eqnarray}
{\cal L}_{\chi ,T}^{(4)} &=&-\frac 1{m^2}\left( J_2^{(2)}:J_2^{(2)}\right)\,,
\nonumber \\
{\cal L}_{\chi ,T}^{(6)} &=&-\frac 2{m^2}\left( J_2^{(2)}:J_2^{(4)}\right)
+\frac 2{m^4}\left( D\cdot J_2^{(2)}\cdot J_2^{(2)}\cdot \overleftarrow{D}%
\right) -\frac 2{m^2}\left( D\cdot J_2^{(2)}\cdot J_1\right)  \nonumber \\
&&+\frac 4{m^4}\left( D\cdot J_2^{(2)}\cdot J_3:J_2^{(2)}\right) +\frac
4{m^4}\left( J_2^{(2)}:J_4:J_2^{(2)}\right) +\frac 4{m^4}\left(
J_2^{(2)}:J_5\cdot D:J_2^{(2)}\right)  \nonumber \\
&&-\frac 8{m^6}(J_2^{(2)}:J_6::J_2^{(2)}J_2^{(2)})\,.  \label{ChiT}
\end{eqnarray}
We can compare eqs.~(\ref{ChiT}) with the corresponding Lagrangian ${\cal
L}_{\chi ,V}^{(6)}$ (\ref{L6V}). One can immediately see that within the
antisymmetric tensor field formalism much richer structure of effective chiral
Lagrangian is obtained. Especially it covers all the structure of the vector
case, with the only exception:
\begin{equation}\label{J1J1}
{\cal L}_{\chi, T} ^{J_1}=-\frac12 J_1\cdot J_1\,,
\end{equation}
which corresponds to the first term of~(\ref{L6V}). We will return to this term
at the beginning of Section~\ref{FO}.

As shown in \cite{CEEKPP}, further simplification of the
Lagrangian (\ref{L_T}) is possible. Provided we are interested only in the
resonance contributions to the low energy constants of the $O(p^6)$ chiral
Lagrangian, the resonance fields take merely a role of the integration
variables and can be therefore freely redefined. For example using {\em linear}
redefinition
\begin{equation}
R\rightarrow R-\frac 2{m^2}\delta J_4:R\,,
\end{equation}
where $\delta J_4=O(p^2)$, we get
\begin{equation}
{\cal L}_T^{(4)}\rightarrow {\cal L}_T^{(4)}-(R:\delta J_4:R)-\frac
2{m^2}\left( J_2^{(2)}:\delta J_4:R\right) +\frac 1{m^2}(R:\delta J_4:\delta
J_4:R)\,.
\end{equation}
The last term is of the order $O(p^8)$, so that it does not contribute to
the $O(p^6)$ chiral Lagrangian after integrating out the resonance fields
and can be therefore dropped. With the same transformation ${\cal L}_T^{(6)}$
is reproduced up to the $O(p^8)$ terms, which can be omitted for the same
reason. Effectively up to the order $O(p^6)$ we can therefore either
eliminate in this way the terms of the form $(R:\delta J_4:R)$ (including
complete elimination of $J_4$) at the price of redefinition of the source
$J_2^{(4)}$
\begin{equation}
J_2^{(4)}\rightarrow J_2^{(4)}-\frac 2{m^2}J_2^{(2)}:\delta J_4\,,
\end{equation}
or eliminate terms which can be written in the form $\left( J_2^{(2)}:\delta
J_4:R\right) $ for suitable $\delta J_4$ which modifies $J_4$ according to
the prescription
\begin{equation}
J_4\rightarrow J_4-\delta J_4\,.
\end{equation}
Namely this second possibility was used in \cite{CEEKPP}. In the same
spirit, using {\em nonlinear} field redefinition
\begin{equation}
R\rightarrow R-\frac 2{m^2}\delta J_6::RR\,,  \label{RforRR}
\end{equation}
where $\delta J_6=O(p^0)$ we get
\begin{eqnarray}
{\cal L}_T^{(4)} &\rightarrow &{\cal L}_T^{(4)}-\left( R:\delta
J_6::RR\right) -\frac 2{m^2}\left( J_2^{(2)}:\delta J_6::RR\right) +O(p^8)\,, \nonumber \\
{\cal L}_T^{(6)} &\rightarrow &{\cal L}_T^{(6)}+O(p^8)\,,
\end{eqnarray}
which effectively means
\begin{eqnarray}
J_4 &\rightarrow &J_4-\frac 2{m^2}J_2^{(2)}:\delta J_6\,,  \nonumber \\
J_6 &\rightarrow &J_6-\delta J_6\,,  \label{J4J6redefRR}
\end{eqnarray}
allowing either elimination of the terms $(R:\delta J_6::RR)$ (including
complete elimination of $J_6$) or cancellation of the terms which can be
expressed as $\left( J_2^{(2)}:\delta J_6::RR\right) $, as it was done in
\cite{CEEKPP}. Of course, all these formal manipulations as well as the
possibility to get rid of $J_1$ and $J_3$ by means of the integration by
parts (\ref{byparts}) are already encoded in (\ref{ChiT}), which can be
rewritten {\em e.g.} in the form
\begin{eqnarray}
{\cal L}_{\chi ,T}^{(4)} &=&-\frac 1{m^2}\left( J_2^{(2)}:J_2^{(2)}\right)\,, \nonumber\\
{\cal L}_{\chi ,T}^{(6)} &=&-\frac 2{m^2}\left( J_2^{(2)}:\widetilde{J}_2^{(4)}\right)
+\frac 2{m^4}\left( D\cdot J_2^{(2)}\cdot J_2^{(2)}\cdot
\overleftarrow{D}\right) +\frac 4{m^4}\left( J_2^{(2)}:\widetilde{J}_5\cdot
D:J_2^{(2)}\right)\,,
\end{eqnarray}
where
\begin{eqnarray*}
\widetilde{J}_2^{(4)} &=&J_2^{(4)}-\frac 2{m^2}J_2^{(2)}:[J_4-\frac
2{m^2}J_2^{(2)}:J_6-\frac 12\widehat{DJ}_3]-\frac 12\widehat{DJ}_1\,, \\
\widetilde{J}_5 &=&J_5-\frac 12\widehat{gJ}_3\,.
\end{eqnarray*}
Lagrangians ${\cal L}_T$ of eq. (\ref{L_T}) and $\widetilde{{\cal L}}_T$,
where
\begin{equation}
\widetilde{{\cal L}}_T =-\frac 12(W\cdot W)+\frac 14m^2(R:R)
+(J_2^{(2)}:R)+(\widetilde{J}_2^{(4)}:R)+(R:\widetilde{J}_5\cdot D:R)\,,
\end{equation}
are not {\em completely} equivalent on the resonance level. However, ${\cal L%
}_T$ and $\widetilde{{\cal L}_T}$ generate the same effective $O(p^6)$
chiral Lagrangian and are therefore {\em strongly }equivalent on the level
of chiral Lagrangians. Though the corresponding Green functions and other
observables at the resonance level are generally different, they
coincide after the chiral expansion up to the order $O(p^6)$ is performed.

\section{Equivalence of the vector and antisymmetric tensor field Lagrangians}\label{S3}

In order to simplify the following discussion, let us consider here a toy
example of one vector and one antisymmetric tensor field without any group
structure. The general situation will be described in the next two
subsections.

As it was recognized in \cite{ecker2}, the naive correspondence connecting
free vector and antisymmetric tensor fields
\begin{eqnarray}
R &\leftrightarrow &\frac 1m\widehat{V}=\frac 1m\widehat{\partial V}\,,
\nonumber \\
V &\leftrightarrow &-\frac 1mW=-\frac 1m\partial \cdot R  \label{RtoV_naive}
\end{eqnarray}
does not relate the Lagrangians (\ref{L_V}) and (\ref{L_T}) properly. The
reason is the difference between the free antisymmetric tensor field
propagator
\begin{eqnarray}
{\rm i}\Delta _F^R(p)_{\mu \nu \,\,\rho \sigma } &=&\int {\rm d}^4x{\rm e}^{%
{\rm i}p\cdot x}\langle 0|T\left( R_{\mu \nu }(x)R_{\rho \sigma }(0)\right)
|0\rangle \\
&=&-\frac{{\rm i}}{p^2-m^2+{\rm i}0}\frac 1{m^2}\left( (m^2-p^2)g_{\mu \rho
}g_{\nu \sigma }+g_{\mu \rho }p_\nu p_\sigma -g_{\mu \sigma }p_\nu p_\rho
-(\mu \leftrightarrow \nu )\right)\nonumber
\end{eqnarray}
and the propagator of $\frac 1m\widehat{V}_{\mu \nu }=\frac 1m(\partial _\mu
V_\nu -\partial _\nu V_\mu )$ in the vector field formalism
\begin{eqnarray}
{\rm i}\Delta _F^{\widehat{V}/m}(p)_{\mu \nu \,\,\rho \sigma } &=&\frac
1{m^2}\int {\rm d}^4x{\rm e}^{{\rm i}p\cdot x}\langle 0|T\left( \widehat{V}%
_{\mu \nu }(x)\widehat{V}_{\rho \sigma }(0)\right) |0\rangle \nonumber\\
&=&\frac 1{m^2}\left( p_\nu p_\sigma {\rm i}\Delta _F^V(p)_{\mu \rho }-p_\nu
p_\rho {\rm i}\Delta _F^V(p)_{\mu \sigma }-(\mu \leftrightarrow \nu )\right)
\nonumber\\
&=&{\rm i}\Delta _F^R(p)_{\mu \nu ,\rho \sigma }-\frac{{\rm i}}{m^2}\left(
g_{\mu \rho }g_{\nu \sigma }-g_{\nu \rho }g_{\mu \sigma }\right)\,,
\end{eqnarray}
where
\begin{eqnarray*}
{\rm i}\Delta _F^V(p)_{\mu \nu } &=&\int {\rm d}^4x{\rm e}^{{\rm i}p\cdot
x}\langle 0|T\left( V_\mu (x)V_\nu (0)\right) |0\rangle \\
&=&-\frac{{\rm i}}{p^2-m^2+{\rm i}0}\left( g_{\mu \nu }-\frac{p_\mu p_\nu }{%
m^2}\right)
\end{eqnarray*}
is the free vector field propagator. Analogically, there is a difference
between the propagator ${\rm i}\Delta _F^V(p)_{\mu \nu }$ and propagator of
$-\frac 1mW_\mu =-\frac 1m\partial ^\nu R_{\nu \mu }$ in the antisymmetric
tensor field formalism
\begin{eqnarray}
{\rm i}\Delta _F^{-W/m}(p)_{\mu \nu } &=&\frac 1{m^2}\int {\rm d}^4x{\rm e}^{%
{\rm i}p\cdot x}\langle 0|T\left( W_\mu (x)W_\nu (0)\right) |0\rangle \nonumber\\
&=&\frac 1{m^2}p^\rho p^\sigma {\rm i}\Delta _F^R(p)_{\rho \mu ,\sigma \nu }
\nonumber\\
&=&{\rm i}\Delta _F^V(p)_{\mu \nu }-\frac{{\rm i}}{m^2}g_{\nu \sigma }\,.
\end{eqnarray}
In both cases, the difference is represented by a {\em contact} term.
Therefore, when passing {\em e.g.} from the tensor to the vector formalism,
the naive substitution (\ref{RtoV_naive}) in the interaction terms must be
supplemented with propagator corrections to each graph with resonance
internal lines. Such corrections correspond to the shrinking of one
(or more) resonance internal line(s) to a point and multiplying by appropriate
power of {\rm i}$/m^2$. Provided we have started with Lagrangian containing
only {\em linear} resonance interaction terms, {\em e.g.}
\begin{equation}
{\cal L}_T=-\frac 12(W\cdot W)+\frac 14m^2(R:R)+(J_2:R)\,,
\end{equation}
the naive correspondence means
\begin{equation}
{\cal L}_T\rightarrow {\cal L}_V=-\frac 14(\widehat{V}:\widehat{V})+\frac
12m^2(V\cdot V)+\frac 1m(J_2:\widehat{V})
\end{equation}
and the only graphs that need the propagator correction are those with one
resonance internal line. Such a correction can be established by adding to
${\cal L}_V$ a contact term \cite{ecker2}
\begin{equation}
{\cal L}_V\rightarrow {\cal L}_V-\frac 1{m^2}\left(
J_2^{(2)}:J_2^{(2)}\right)
\end{equation}
and the resulting Lagrangian is then {\em completely} equivalent to ${\cal L}%
_T$. The {\em bi(tri)linear} couplings, however, generate infinite number of
graphs, which should be corrected. This leads to the necessity to add an
infinite number of additional contact terms on the lagrangian level to
ensure the complete equivalence. Finding all such terms in a systematic way,
as well as discussion of their relevance for the effective $O(p^6)$ chiral
Lagrangian is the subject of the next two subsections.

\subsection{Vector $\rightarrow$ tensor correspondence}

In this subsection we shall start with the vector field Lagrangian ${\cal L}%
_V$ given by eq. (\ref{L_V}) and try to construct antisymmetric tensor field
Lagrangian ${\cal L}_T^{\eff}$ which is {\em completely} equivalent to ${\cal %
L}_V$ on the resonance level. As we have mentioned above, such a Lagrangian
will contain infinite number of terms with increasing chiral order. Provided
we claim to get a local Lagrangian with finite number of terms up to the
order $O(p^6)$, we obtain only {\em incomplete} equivalence on the resonance
level (however {\em strong} equivalence on the level of the chiral Lagrangian).

Let us consider the (generally nonlocal) (pseudo)Goldstone boson effective
action $\Gamma _V[J_i,K]$ defined as
\begin{equation}
Z_V[J_i,K]=\exp {\rm i}\Gamma _V[J_i,K]=\int {\cal D}V\exp \left( {\rm i}%
\int {\rm d}^4x{\cal L}_V\right)\,.
\end{equation}
The vector resonance contribution to the effective chiral Lagrangian is then
obtained by the expansion of $\Gamma _V[J_i,K]$ up to the order $O(p^6)$.
Complete equivalence of ${\cal L}_V$ and ${\cal L}_T^{\eff}$ means
\begin{equation}
Z_V[J_i,K]=\exp {\rm i}\Gamma _V[J_i,K]=\int {\cal D}R\exp \left( {\rm i}%
\int {\rm d}^4x{\cal L}_T^{\eff}\right)\,.
\end{equation}
In what follows we use the method of integrating in an additional field, cf.
\cite{pallante}, \cite{bijnens} and \cite{tanabashi}. Introducing an
auxiliary antisymmetric tensor field $R$, we can write
\begin{equation}
Z_V[J_i,K]=\int {\cal D}V\exp \left( {\rm i}\int {\rm d}^4x{\cal L}_V\right)
=\frac{\int {\cal D}V{\cal D}R\exp \left( {\rm i}\int {\rm d}^4x(\frac
14(R:R)+{\cal L}_V)\right) }{\int {\cal D}R\exp \left( {\rm i}\int {\rm d}%
^4x(\frac 14(R:R)\right) }\,.
\end{equation}
The auxiliary field $R$ is merely an integration variable, so it can be
freely redefined. We can use this freedom in order to get rid of
the terms involving derivatives of the field $V$ and simplify in this way
the $V$ integration. The desired redefinition corresponds to a shift
according to the prescription
\begin{equation}
R\rightarrow mR-\widehat{V}+2J_2+2V\cdot J_3
\end{equation}
and as a result we get
\begin{equation}
Z_V[J_i,K]=\frac{\int {\cal D}V{\cal D}R\exp \left( {\rm i}\int {\rm d}^4x%
{\cal L}_{VR}\right) }{\int {\cal D}R\exp \left( {\rm i}\int {\rm d}%
^4x(\frac 14m^2(R:R)\right) }\,.
\end{equation}
Here ${\cal L}_{VR}$ (up to a total derivative) can be written as
\begin{equation}
{\cal L}_{VR}={\cal L}_T^{\prime }+\frac 12m^2\left( V\cdot {\cal K}\cdot
V\right) +\left( {\cal J}\cdot V\right)\,,
\end{equation}
where
\begin{eqnarray*}
{\cal L}_T^{\prime } &=&\frac 14m^2\left( R:R\right) +m\left( J_2:R\right)
+\left( J_2:J_2\right)\,,  \\
{\cal K} &=&1+\frac K{m^2}+\frac 2{m^2}J_3:J_3\,, \\
{\cal J} &=&J_1+mR:J_3+2J_2:J_3+mW\,,
\end{eqnarray*}
with $W=D\cdot R$, (cf. (\ref{W})). The integration over vector field is now
straightforward and we have
\begin{equation}
Z_V[J_i,K]=\frac{\int {\cal D}R\exp \left( -\frac{{\rm i}}2Tr\ln \left(
\frac{m^2{\cal K}}{2\pi }\right) +{\rm i}\int {\rm d}^4x\left( {\cal L}%
_T^{\prime }-\frac 1{2m^2}{\cal J\cdot K}^{-1}\cdot {\cal J}\right) \right)
}{\int {\cal D}R\exp \left( {\rm i}\int {\rm d}^4x(\frac 14m^2(R:R)\right) }\,.
\end{equation}
Because the kernel ${\cal K}$ is local, $Tr\ln \left( \frac{m^2{\cal K}}{%
2\pi }\right) \propto \delta ^{(4)}(0)$ and can be dropped within
dimensional regularization. Writing further
\begin{equation}
{\cal K}^{-1}=1+\sum_{n=1}^\infty \left( -\frac 1{m^2}\right)
^n(K+2J_3:J_3)^n\,,
\end{equation}
we get
\begin{equation}
Z_V[J_i,K]=\frac{\int {\cal D}R\exp \left( {\rm i}\int {\rm d}^4x{\cal L}%
_T^{\eff}\right) }{\int {\cal D}R\exp \left( {\rm i}\int {\rm d}^4x(\frac
14m^2(R:R)\right) }\,.
\end{equation}
with Lagrangian ${\cal L}_T^{\eff}$ (counting now $R=O(p^2)$ as usual)
\begin{equation}\label{L_T_eff_complet}
{\cal L}_T^{\eff}={\cal L}_T^{\eff(4)}+{\cal L}_T^{\eff(6)}+\sum_{n=1}^\infty {\cal L}_T^{\eff(2n+6)}\,,
\end{equation}
where
\begin{eqnarray*}
{\cal L}_T^{\eff(4)} &=&\frac 14m^2\left( R:R\right) +m\left( J_2:R\right)
+\left( J_2:J_2\right)\,,\\
{\cal L}_T^{\eff(6)} &=&-\frac 12\left( W\cdot W\right) -\left( W\cdot
J_3:R\right) -\frac 12\left( R:J_3\cdot J_3:R\right)\\
&&-\frac 1m\left( J_1\cdot W\right) -\frac 2m\left( J_2:J_3\cdot W\right)
-\frac 2m\left( J_2:J_3\cdot J_3:R\right) -\frac 1m\left( J_1\cdot
J_3:R\right)\\
&&-\frac 2{m^2}\left( J_2:J_3\cdot J_1\right) -\frac 1{2m^2}\left( J_1\cdot
J_1\right) -\frac 2{m^2}\left( J_2:J_3\cdot J_3:J_2\right)\,,\\
{\cal L}_T^{\eff(2n+6)} &=&\frac 12\left( -\frac 1{m^2}\right) ^{n+1}\left(
J_1+mR:J_3+2J_2:J_3+mW\right)\\
&&\cdot \left( K+2J_3:J_3\right) ^n\cdot \left(
J_1+mR:J_3+2J_2:J_3+mW\right)\,.
\end{eqnarray*}
The Lagrangian ${\cal L}_T^{\eff}$ is completely equivalent to ${\cal L}_V$
(given by (\ref{L_V})) at the resonance level by construction. Note that,
infinite number of terms (including infinite number of contact terms) is
necessary to ensure this complete equivalence unless $K=J_3=0$ in the
original model.

The antisymmetric tensor field Lagrangian ${\cal L}_T^{\eff,(\leq 6)}$ of the
form (\ref{L_T}) which is {\em weakly} equivalent up to the order $O(p^6)$
to original ${\cal L}_V$ can be obtained by truncation of the infinite
series and omitting the contact terms
\begin{align}\label{Leffm6}
{\cal L}_T^{\eff(\leq 6)} &=-\frac 12\left( W\cdot W\right) +\frac
14m^2\left( R:R\right) \\
&+\left( J_1^{\eff}\cdot W\right) +\left( J_2^{\eff}:R\right) +\left( W\cdot
J_3^{\eff}:R\right) +\left( R:J_4^{\eff}:R\right) +\left( R:J_5^{\eff}\cdot
D:R\right)\,,\notag
\end{align}
where
\begin{eqnarray}
J_1^{\eff} &=&-\frac 1mJ_1-\frac 2mJ_2:J_3\,, \nonumber\\
J_2^{\eff} &=&mJ_2-\frac 2mJ_2:J_3\cdot J_3-\frac 1mJ_1\cdot J_3\,, \nonumber\\
J_3^{\eff} &=&-J_3\,, \nonumber\\
J_4^{\eff} &=&-\frac 12J_3\cdot J_3\,, \nonumber\\
J_5^{\eff} &=&0\,.\label{Jieff}
\end{eqnarray}
The additional contact terms ${\cal L}_T^{\eff,\,(\leq 6)contact}$, which are
necessary to get {\em strong} equivalence are
\begin{equation}\label{Leffm6c}
{\cal L}_T^{\eff,\,(\leq 6)contact}=\left( J_2:J_2\right) -\frac
1{2m^2}\left( J_1\cdot J_1\right) -\frac 2{m^2}\left( J_2:J_3\cdot
J_1\right) -\frac 2{m^2}\left( J_2:J_3\cdot J_3:J_2\right)\, .
\end{equation}

In the next subsection we will deal with tensor Lagrangian and we will
`transform' it to the vector one. We can formally do it already now by means of
inversion of relations (\ref{Jieff}) and substitution into (\ref{L_V}) and
adding transformed (\ref{Leffm6c}) with opposite sign. However, as we shall see
this does not produce the complete equivalence.

\subsection{Tensor $\rightarrow$ vector correspondence}

In this subsection we reverse the consideration of the previous subsection and
try to construct vector Lagrangian ${\cal L}_V^{\eff}$ which is equivalent to
tensor one ${\cal L}_T$ (see (\ref{L_T})) on the resonance level. For the
technical reason (and for simplification of the resulting formulae), we first
perform integration by parts to include $J_3$ into $J_4$ and $J_5$ (cf.
(\ref{byparts})) and also a nonlinear field
redefinition (\ref{RforRR}) in order to get rid of the cubic terms in ${\cal %
L}_T$ (at the price that we lose of course the {\em complete} equivalence
already at this stage). That means we start with the Lagrangian
\begin{eqnarray}
{\cal L}_T &=&-\frac 12(W\cdot W)+\frac 14m^2(R:R)  \nonumber \\
&&+(J_1\cdot W)+(J_2:R)+(R:J_4:R)+(R:J_5\cdot D:R)\,. \label{start_L_T}
\end{eqnarray}
The next steps are then the same as in the previous case. Let us define the
nonlocal (pseudo)Goldstone boson effective action $\Gamma _T[J_i]$ as
\begin{equation}
Z_T[J_i]=\exp {\rm i}\Gamma _T[J_i]=\int {\cal D}R\exp \left( {\rm i}\int
{\rm d}^4x{\cal L}_T\right)
\end{equation}
and introduce an auxiliary vector field $V$%
\begin{equation}
Z_T[J_i]=\frac{\int {\cal D}V{\cal D}R\exp \left( {\rm i}\int {\rm d}%
^4x(\frac 12(V\cdot V)+{\cal L}_T)\right) }{\int {\cal D}V\exp \left( {\rm i}%
\int {\rm d}^4x\frac 12(V\cdot V)\right) }\,.
\end{equation}
After a field redefinition, designed in order to cancel the kinetic term of
the $R$ field (and other derivative terms)
\begin{equation}
V\rightarrow mV-W+J_1\,,
\end{equation}
we get
\begin{equation}
Z_T[J_i]=\frac{\int {\cal D}V{\cal D}R\exp \left( {\rm i}\int {\rm d}^4x%
{\cal L}_{RV}\right) }{\int {\cal D}V\exp \left( {\rm i}\int {\rm d}%
^4x(\frac 12m^2(V\cdot V)\right) }\,.
\end{equation}
After integration by parts and up to a total derivative we can write
\begin{equation}\label{L_RV}
{\cal L}_{RV}={\cal L}_V^{^{\prime }}+\frac 14m^2\left( R:{\cal K}:R\right)
+\left( R:{\cal J}\right)\,,
\end{equation}
where now
\begin{eqnarray*}
{\cal L}_V^{^{\prime }} &=&\frac 12m^2(V\cdot V)+m\left( J_1\cdot V\right)
+\frac 12\left( J_1\cdot J_1\right)\,,  \\
{\cal K} &=&1+\frac 4{m^2}J_4+\frac 4{m^2}J_5\cdot D\,, \\
{\cal J} &=&J_2+\frac 12m\widehat{V}\,.
\end{eqnarray*}
Because we effectively (up to the order $O(p^6)$) incorporated the cubic
term into the source $J_4$ (cf.(\ref{J4J6redefRR})), the integration over $R$
is gaussian and we obtain
\begin{equation}
Z_T[J_i]=\frac{\int {\cal D}V\exp \left( -\frac{{\rm i}}2Tr\ln \left( \frac{%
m^2{\cal K}}{4\pi }\right) +{\rm i}\int {\rm d}^4x\left( {\cal L}%
_V^{^{\prime }}-\frac 1{m^2}\left( {\cal J}:{\cal K}^{-1}:{\cal J}\right)
\right) \right) }{\int {\cal D}V\exp \left( {\rm i}\int {\rm d}^4x(\frac
12m^2(V\cdot V)\right) }\,.
\end{equation}
Within dimensional regularization, we can again drop the term $Tr\ln \left(
\frac{m^2{\cal K}}{4\pi }\right) $, the chiral expansion of which is now
proportional to the derivatives of delta function at zero argument. Writing
further
\begin{equation}
{\cal K}^{-1}=1+\sum_{n=1}^\infty (-1)^n\left( \frac 2m\right) ^{2n}\left(
J_4+J_5\cdot D\right)^n\,,
\end{equation}
we get
\begin{equation}
Z_T[J_i]=\frac{\int {\cal D}V\exp \left( {\rm i}\int {\rm d}^4x{\cal L}%
_V^{\eff}\right) }{\int {\cal D}V\exp \left( {\rm i}\int {\rm d}^4x(\frac
12m^2(V\cdot V)\right) }\,,
\end{equation}
where
\begin{eqnarray}
{\cal L}_V^{\eff} &=&\frac 12m^2(V\cdot V)+m\left( J_1\cdot V\right) +\frac
12\left( J_1\cdot J_1\right)   \nonumber \\
&&-\frac 14\left( \widehat{V}:\widehat{V}\right) -\frac 1m\left( J_2:%
\widehat{V}\right) -\frac 1{m^2}\left( J_2:J_2\right)   \nonumber \\
&&-\frac 1{m^2}\sum_{n=1}^\infty (-1)^n\left( \frac 2m\right) ^{2n}\left(
{\cal J}:\left( J_4+J_5\cdot D\right) ^n:{\cal J}\right)\,.
\label{L_Veff_Complete}
\end{eqnarray}
Counting $V=O(p^3)$ as usual, we have
\begin{equation}
{\cal L}_V^{\eff}={\cal L}_V^{\eff(4)}+{\cal L}_V^{\eff(6)}+{\cal L}_V^{\eff(8)}+{\cal L}%
_V^{\eff(>8)}\,,
\end{equation}
where, after splitting $J_2=J_2^{(2)}+J_2^{(4)}=O(p^2)+O(p^4)$%
\begin{eqnarray}
{\cal L}_V^{\eff(4)} &=&-\frac 1{m^2}\left( J_2^{(2)}:J_2^{(2)}\right)\,,
\nonumber \\
{\cal L}_V^{\eff(6)} &=&\frac 12m^2(V\cdot V)+m\left( J_1\cdot V\right) -\frac
1m\left( J_2^{(2)}:\widehat{V}\right) -\frac 2{m^2}\left(
J_2^{(2)}:J_2^{(4)}\right)   \nonumber \\
&&+\frac 12\left( J_1\cdot J_1\right) +\frac 4{m^4}\left(
J_2^{(2)}:J_4:J_2^{(2)}\right) +\frac 4{m^4}\left( J_2^{(2)}:J_5\cdot
D:J_2^{(2)}\right)\,,   \nonumber \\
{\cal L}_V^{\eff(8)} &=&-\frac 14\left( \widehat{V}:\widehat{V}\right) -\frac
1m\left( J_2^{(4)}:\widehat{V}\right) +\frac 4{m^3}\left( J_2^{(2)}:J_4:%
\widehat{V}\right)   \nonumber \\
&&+\frac 2{m^3}\left( J_2^{(2)}:J_5\cdot D:\widehat{V}\right) +\frac
2{m^3}\left( \widehat{V}:J_5\cdot D:J_2^{(2)}\right)   \nonumber \\
&&-\frac 1{m^2}\left( J_2^{(4)}:J_2^{(4)}\right) +\frac 8{m^4}\left(
J_2^{(2)}:J_4:J_2^{(4)}\right)   \nonumber \\
&&+\frac 4{m^4}\left( J_2^{(4)}:J_5\cdot D:J_2^{(2)}\right) +\frac
4{m^4}\left( J_2^{(2)}:J_5\cdot D:J_2^{(4)}\right)   \nonumber \\
&&-\frac{16}{m^6}\left( J_2^{(2)}:\left( J_4+J_5\cdot D\right)
^2:J_2^{(2)}\right)\,.   \label{L_Veff_4_6_8}
\end{eqnarray}
A complete form of ${\cal L}_V^{\eff(>8)}$, which starts from $O(p^{10})$ can be
easily deduced from (\ref{L_Veff_Complete}). In the next section we will
need only the explicit form of the terms of the order $O(p^{10})$%
\begin{align}
{\cal L}_V^{\eff(10)} =&\phantom{+}\frac 4{m^4}\left( J_2^{(4)}:J_4:J_2^{(4)}\right)
+\frac 4{m^3}\left( J_2^{(4)}:J_4:\widehat{V}\right) +\frac 1{m^2}\left(
\widehat{V}:J_4:\widehat{V}\right) +\frac 1{m^2}\left( \widehat{V}:J_5\cdot
D:\widehat{V}\right)   \notag \\
&+\frac 2{m^3}\left( J_2^{(4)}:J_5\cdot D:\widehat{V}\right) +\frac
2{m^3}\left( \widehat{V}:J_5\cdot D:J_2^{(4)}\right) +\frac 4{m^4}\left(
J_2^{(4)}:J_5\cdot D:J_2^{(4)}\right)   \notag \\
&-\frac{16}{m^6}\left( J_2^{(4)}:\left( J_4+J_5\cdot D\right)
^2:J_2^{(2)}\right) -\frac{16}{m^6}\left( J_2^{(2)}:\left( J_4+J_5\cdot
D\right) ^2:J_2^{(4)}\right)   \notag \\
&-\frac 8{m^5}\left( \widehat{V}:\left( J_4+J_5\cdot D\right)
^2:J_2^{(2)}\right) -\frac 8{m^5}\left( J_2^{(2)}:\left( J_4+J_5\cdot
D\right) ^2:\widehat{V}\right)   \notag \\
&+\frac{64}{m^8}\left( J_2^{(2)}:\left( J_4+J_5\cdot D\right)
^3:J_2^{(2)}\right)\,.   \label{L_Veff_10}
\end{align}
Collecting terms up to the order $O(p^6)$ we get {\em weakly} equivalent vector
Lagrangian of the form (\ref{L_V}) corresponding to the tensor one
(\ref{start_L_T})\footnote{Here we have also to include the kinetic term which is formally of the order $O(p^8)$.}
\begin{eqnarray}
{\cal L}_V^{\eff(\leq 6)} &=&-\frac 14\left( \widehat{V}:\widehat{V}\right)
+\frac 12m^2(V\cdot V) \nonumber\\
&&+(J_1^{\eff}\cdot V)+\left( J_2^{\eff}:\widehat{V}\right) +\frac 12(V\cdot
K^{\eff}\cdot V)+\left( V\cdot J_3^{\eff}:\widehat{V}\right)
\end{eqnarray}
with
\begin{eqnarray*}
J_1^{\eff} &=&mJ_1\,, \\
J_2^{\eff} &=&-\frac 1mJ_2^{(2)}\,, \\
K^{\eff} &=&J_3^{\eff}=0\,.
\end{eqnarray*}
In order to ensure {\em strong} equivalence, we have to add $O(p^6)$ contact
terms of the form
\begin{eqnarray}
{\cal L}_V^{\eff(\leq 6),contact} &=&\frac 12\left( J_1\cdot J_1\right)
-\frac 1{m^2}\left( J_2^{(2)}:J_2^{(2)}\right) -\frac 2{m^2}\left(
J_2^{(2)}:J_2^{(4)}\right)  \nonumber\\
&&+\frac 4{m^4}\left( J_2^{(2)}:J_4:J_2^{(2)}\right) +\frac 4{m^4}\left(
J_2^{(2)}:J_5\cdot D:J_2^{(2)}\right) .
\end{eqnarray}
Provided we had started with the more rich form of ${\cal L}_T$ (cf. (\ref
{L_T})), we should insert
\begin{eqnarray}
J_2^{(4)} &\rightarrow &J_2^{(4)}+\frac 4{m^4}J_6::J_2^{(2)}J_2^{(2)}\,, \nonumber\\
J_4 &\rightarrow &J_4-\frac 12\widehat{DJ}_3\,, \nonumber\\
J_5 &\rightarrow &J_5-\frac 12\widehat{gJ}_3
\end{eqnarray}
in the above expressions.

As in the previous case, we cannot achieve complete equivalence with finite
number of terms (even if the cubic interaction of the resonance field is
missing from the very beginning), unless $J_3=J_4=J_5=0$.

\section{Explicit example of complete versus strong equivalence - $VVP$
correlator}\label{S4}

In this section we shall illustrate the above general constructions using
the calculation of the $VVP$ correlator in the chiral limit within several
variants of the resonance effective theory. We use the notation of Ref. \cite
{Moussallam95} (cf. \cite{RuizFemenia} and \cite{knecht})
\begin{equation}
\int {\rm d}^4x{\rm d}^4y{\rm e}^{{\rm i}(p\cdot x+q\cdot y)}\langle
0|T\left( V_\mu ^a(x)V_\nu ^b(y)P^c(0)\right) |0\rangle =\varepsilon _{\mu
\nu \alpha \beta }p^\alpha q^\beta d^{abc}\Pi _{VVP}(p^2,q^2,r^2)\,,
\label{VVP_def}
\end{equation}
where $r=-(p+q)$,
\begin{eqnarray*}
V_\mu ^a &=&\frac 1{\sqrt{2}}\overline{q}\gamma _\mu T^aq\,, \\
P^a &=&\frac 1{\sqrt{2}}\overline{q}i\gamma _5T^aq
\end{eqnarray*}
and $T^a=\lambda ^a/\sqrt{2}$. The right hand side of (\ref{VVP_def})
follows from invariance with respect to $SU(3)_V$, $P$ and $T$
transformations as well as from chiral Ward identities. In the low energy
limit, the behaviour of $\Pi _{VVP}(p^2,q^2,r^2)$ is governed by chiral
perturbation theory, the corresponding low energy expansion up to the order
$O(p^6)$ reads \cite{knecht}
\begin{equation}
\Pi _{VVP}^{\chi PT}(p^2,q^2,r^2)=\frac{B_0}{r^2}\left( -\frac{N_C}{8\pi ^2}%
+4(A_2-4A_3)(p^2+q^2)+4(-A_2+2A_3+4A_4)r^2+\ldots \right)\,.  \label{VVP_CHPT}
\end{equation}
The first term in the brackets is fixed by the chiral anomaly and
corresponds to the contribution of the $O(p^4)$ Wess-Zumino term \cite{Wess, Gasser:1984gg}.
The remaining terms are the contributions of the $O(p^6)$ contact terms with odd
intrinsic parity, as listed in the Ref. \cite{Fearing}, and the ellipses
stand for the loop contributions, which are suppressed in the large $N_C$
limit and will not be considered here. In the counterterm basis given in
\cite{BGT}, the formula (\ref{VVP_CHPT}) is rewritten as (cf. also \cite{kkn})
\begin{equation}
\Pi _{VVP}^{\chi PT}(p^2,q^2,r^2)=\frac{B_0}{r^2}\left( -\frac{N_C}{8\pi ^2}%
+4C^W_{13}(p^2+q^2)-8(C^W_{11}-4C^W_3+4C^W_7)r^2+\ldots \right)\,. \label{VVP_CHPT2}
\end{equation}

The short distance asymptotics in the case when one or both momenta are
simultaneously large is given by OPE. In the chiral limit and in the leading
order in $\alpha _s$ we have \cite{Moussallam95}, \cite{knecht}
\begin{eqnarray}
\Pi _{VVP}(\lambda ^2p^2,\lambda ^2q^2,\lambda ^2r^2) &=&\frac{B_0F_0^2}{%
2\lambda ^4}\frac{p^2+q^2+r^2}{p^2q^2r^2}+O\left( \frac 1{\lambda ^6}\right)\,,  \nonumber \\
\Pi _{VVP}(\lambda ^2p^2,(q-\lambda p)^2,q^2) &=&\frac{B_0F_0^2}{\lambda ^2}%
\frac 1{p^2q^2}+O\left( \frac 1{\lambda ^3}\right)\,,  \nonumber \\
\Pi _{VVP}(\lambda ^2p^2,q^2,(q+\lambda p)^2) &=&\frac 1{\lambda ^2}\frac
1{p^2}\Pi _{VT}(q^2)+O\left( \frac 1{\lambda ^6}\right)\,.  \label{OPE}
\end{eqnarray}
where $\Pi _{VT}(q^2)$ is defined as
\begin{equation}
\int {\rm d}^4x{\rm e}^{{\rm i}q\cdot x}\langle 0|T\left( V_\mu ^a(x)T_{\rho
\sigma }^b(0)\right) |0\rangle =(q_\rho g_{\mu \sigma }-q_\sigma g_{\mu \rho
})\Pi _{VT}(q^2)\delta ^{ab}\,,
\end{equation}
with
\[
T_{\rho \sigma }^b=\frac 1{\sqrt{2}}\overline{q}\sigma _{\rho \sigma }T^aq\,.
\]
The chiral resonance theory should yield $\Pi _{VVP}$ in the intermediate
energy region.

Let us start with the tensor field formulation using the Lagrangian (\ref
{L_T}) to which we add $O(p^2)$ chiral Lagrangian as well as the Wess-Zumino
term. The relevant sources $J_i$ in (\ref{L_T}) potentially contributing to
$VVP$ three-point function are listed in Appendix \ref{apJT}.
The result of the calculation corresponding to the Feynman graphs depicted
in the Fig.~\ref{fig1} is \cite{RuizFemenia}
\begin{multline}
\Pi _{VVP}^T(p^2,q^2,r^2) =B_0\left[ -\frac{N_C}{16\pi^2r^2}+ 2F_V^2
\frac{(d_1+8 d_2 -d_3)r^2+2 d_3 p^2}{(p^2-m^2)(q^2-m^2)r^2}\right.\\
\left. +2\sqrt{2}\frac{F_V}{m}
\frac{(c_1+c_2+8 c_3 -c_5)r^2+(-c_1+c_2+c_5-2c_6)p^2+(c_1-c_2+c_5)q^2}{(p^2-m^2)r^2}
+ (p\leftrightarrow q)\right]\,.\label{VVP_T}
\end{multline}
\begin{figure}[tb]
\begin{center}
\epsfig{file=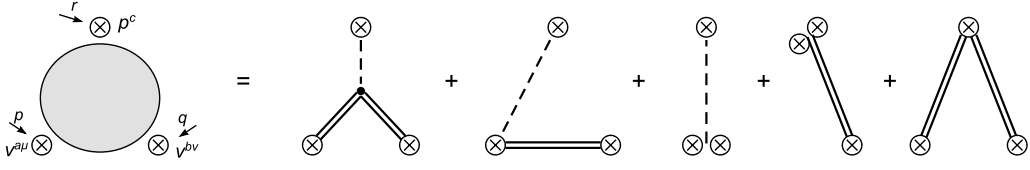}
\end{center}
\caption{Diagram representation of the $VVP$ three point function in tensor field
formalism. Double lines stand for vector resonances in the given formalism
and dash lines for pseudoscalar mesons. Here and in the following figures, crossing is tacitly assumed.}
\label{fig1}
\end{figure}
This expression satisfies the short distance constraints (\ref{OPE}) provided
\cite{RuizFemenia}
\begin{eqnarray}
c_1 &=&-4c_3\,, \nonumber\\
c_2 &=&-4c_3+c_5\,,\nonumber \\
c_6 &=&c_5-\frac{N_C}{64\sqrt{2}\pi ^2}\frac m{F_V}\,, \nonumber\\
d_1 &=&-8d_2-\frac{N_C}{64\pi ^2}\frac{m^2}{F_V^2}+\frac{F_0^2}{4F_V^2}\,, \nonumber\\
d_3 &=&-\frac{N_C}{64\pi ^2}\frac{m^2}{F_V^2}+\frac{F_0^2}{8F_V^2}\,.
\end{eqnarray}
In this case eq.~(\ref{VVP_T}) simplifies to
\begin{equation}
\Pi _{VVP}^T(p^2,q^2,r^2)=\frac{B_0F_0^2}2\frac{p^2+q^2+r^2-\frac{N_C}{4\pi
^2}\frac{m^4}{F_0^2}}{(p^2-m^2)(q^2-m^2)r^2}\,,  \label{LMD}
\end{equation}
which coincides with the lowest meson dominance (LMD) approximation
developed in \cite{Moussallam95}, \cite{knecht}.

Let us compare (\ref{VVP_T}) with the analogical calculation using instead
of ${\cal L}_T$ the strongly equivalent vector field form of the Lagrangian
${\cal L}_V^{\eff(\leq 6)}+{\cal L}_V^{\eff(\leq 6),contact}$. Because we have lost the
complete equivalence by means of the truncation of the infinite series in
(\ref{L_Veff_Complete}), we expect that the result will be generally
different, yielding however the same low energy expansion up to the order
$O(p^6)$. This is of course the case, the result is
\begin{eqnarray}
\Pi _{VVP}^{V \eff}(p^2,q^2,r^2) &=&\frac{B_0}{r^2}\left[ -\frac{N_C}{8\pi ^2}%
+4F_V^2\frac{(d_1-d_3+8d_2)r^2+d_3(p^2+q^2)}{m^4}\right.  \nonumber \\
&&\left. -4\sqrt{2}F_V\frac{(c_1+c_2-c_5+8c_3)r^2+(c_5-c_6)(p^2+q^2)}{m^3}%
\right]\,.  \label{VVP_V_eff}
\end{eqnarray}
It is immediately  seen that $\Pi _{VVP}^{V\eff}(p^2,q^2,r^2)$ does not allow for
the short distance constraint to be satisfied.

A natural question arises here, whether it is possible to recover the full
expression (\ref{VVP_T}) using some extension of the Lagrangian ${\cal L}%
_V^{\eff(\leq 6)}+{\cal L}_V^{\eff(\leq 6),\,contact}$ by means of an addition of only a
{\em finite} number of terms from (\ref{L_Veff_Complete}). The answer is
positive for the following simple reason. Generally, the only relevant
vertices which contribute to the tree graph corresponding to the $n$-point
function are those satisfying $n_S+n_L\leq n$, where $n_S$ is $\,$the number
of the external sources $v$, $a$, $s$ or $p$ and $n_L$ is number of legs.
Because each $J_i$ contains at least one of the external sources or at least
one Goldstone boson leg, only finite number of terms of the Lagrangian (\ref
{L_Veff_Complete}) does really contribute to the $n$-point functions with $n$
fixed. In our example of three-point $VVP$ function, we can safely neglect
all the contact terms composed from more than three $J_i$'s as well as the
resonance interaction terms with one (two) resonance field and more than two
(one) $J_i$'s. Generally, provided we require to reproduce in the vector field
formalism all the three-point functions derived from ${\cal L}_T$, we can
use {\em e.g.} the Lagrangian
\begin{equation}
{\cal L}_V^{\eff,\,{\rm 3-point}\,}={\cal L}_V^{\eff(6)}+{\cal L}%
_V^{\eff(6),\,contact}+{\cal L}_V^{(8)\,{\rm 3-point}}+{\cal L}_V^{(10)\,\,%
{\rm 3-point}}
\end{equation}
with additional $O(p^8)$ and $O(p^{10})$ terms, where (cf. (\ref
{L_Veff_4_6_8}), (\ref{L_Veff_10}))
\begin{eqnarray}
{\cal L}_V^{(8)\,{\rm 3-point}} &=&-\frac 14\left( \widehat{V}:\widehat{V}%
\right) -\frac 1m\left( J_2^{(4)}:\widehat{V}\right) +\frac 4{m^3}\left(
J_2^{(2)}:J_4:\widehat{V}\right)  \nonumber \\
&&+\frac 2{m^3}\left( J_2^{(2)}:J_5\cdot D:\widehat{V}\right) +\frac
2{m^3}\left( \widehat{V}:J_5\cdot D:J_2^{(2)}\right)  \nonumber \\
&&-\frac 1{m^2}\left( J_2^{(4)}:J_2^{(4)}\right) +\frac 8{m^4}\left(
J_2^{(2)}:J_4:J_2^{(4)}\right)  \nonumber \\
&&+\frac 4{m^4}\left( J_2^{(4)}:J_5\cdot D:J_2^{(2)}\right) +\frac
4{m^4}\left( J_2^{(2)}:J_5\cdot D:J_2^{(4)}\right)
\end{eqnarray}
and
\begin{eqnarray}
{\cal L}_V^{(10)\,\,{\rm 3-point}} &=&\frac 4{m^4}\left(
J_2^{(4)}:J_4:J_2^{(4)}\right) +\frac 4{m^3}\left( J_2^{(4)}:J_4:\widehat{V}%
\right) +\frac 1{m^2}\left( \widehat{V}:J_4:\widehat{V}\right)  \nonumber \\
&&+\frac 1{m^2}\left( \widehat{V}:J_5\cdot D:\widehat{V}\right) +\frac
2{m^3}\left( J_2^{(4)}:J_5\cdot D:\widehat{V}\right)  \nonumber \\
&&+\frac 2{m^3}\left( \widehat{V}:J_5\cdot D:J_2^{(4)}\right) +\frac
4{m^4}\left( J_2^{(4)}:J_5\cdot D:J_2^{(4)}\right) \,.
\end{eqnarray}
We have explicitly verified this statement for the $VVP$ three-point function.

For completeness, let us also briefly discuss the reversed problem, {\em i.e.}
let us start with the general vector field Lagrangian ${\cal L}_V$ (see (\ref
{L_V}) and Appendix \ref{apJ}) and compare the result with the strongly equivalent
antisymmetric tensor field one ${\cal L}_T^{\eff,(\leq 6)}+{\cal L}_T^{\eff(\leq 6),\,contact}$.
Using ${\cal L}_V$, the relevant sources $J_i$ in
the vector case are those with the couplings $h_V$, $f_V$ and $\sigma_V$
from Appendix \ref{apJV} and the diagrams are depicted in Fig.~\ref{fig2}.
\begin{figure}[tb]
\begin{center}
\epsfig{file=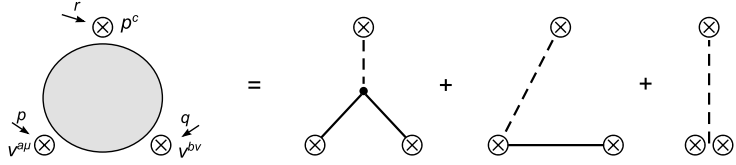}
\end{center}
\caption{Diagram representation of the $VVP$ three point function in vector field
formalism. Vector resonances are depicted now by single lines.}
\label{fig2}
\end{figure}
For the $VVP$ correlator we get the following expression \cite{knecht}
\begin{eqnarray}
\Pi _{VVP}^V(p^2,q^2,r^2) &=&\frac{B_0}{r^2}\left[ -\frac{N_C}{8\pi ^2}%
-4\sigma _Vf_V^2\frac{p^2q^2}{(p^2-m^2)(q^2-m^2)}\right.  \nonumber \\
&&\left. +4\sqrt{2}h_Vf_V\frac{p^2}{(p^2-m^2)}+4\sqrt{2}h_Vf_V\frac{q^2}{%
(q^2-m^2)}\right]\,,  \label{VVP_V}
\end{eqnarray}
which, however, does not satisfy the short distance constraints (\ref{OPE})
even if the $O(p^6)$ contact terms of the form (\ref{VVP_CHPT}), (\ref{VVP_CHPT2}) are added.
On the contrary to the previous case, the Lagrangian ${\cal L}_T^{\eff,(\leq 6)}+%
{\cal L}_T^{\eff(\leq 6),\,contact}$ reproduces this expression completely. The
reason is, that ${\cal L}_T^{\eff,(\leq 6)}+{\cal L}_T^{\eff(\leq
6),\,contact}$ contains already all the necessary terms (note that, ${\cal
L}_T^{\eff(2n+6)}$ with $n\geq 1$ involves only terms with at least four
$J_i$'s and/or
resonance and Goldstone boson legs and ${\cal L}_T^{\eff,(\leq 6)}+{\cal L}%
_T^{\eff(\leq 6),\,contact}$ therefore already recovers all the tree level three
point functions).

\section{First order formalism}
\label{FO}
\subsection{Motivation}

As we have seen in the previous sections, neither the antisymmetric tensor nor
the vector representation is ``complete'' in the sense that when passing to the
$O(p^6)$ effective chiral Lagrangian, some structures possible in one of them
are missing in the other and vice versa. More precisely, as it was mentioned in
\cite{moussallam} and \cite{CEEKPP}, the antisymmetric tensor formalism does not create the contact term of the form
(cf.~(\ref{J1J1}))
\begin{equation}
{\cal L}_\chi ^{J_1}=-\frac 12\left( J_1\cdot J_1\right)\,,
\end{equation}
though the term $(J_1.W)$ is presented in the resonance Lagrangian. In \cite{moussallam}
for instance the following $J_1$ was considered in this context
\begin{equation}
J_1^a=-\frac{f_\chi }m\langle T^a[\chi _{-},u_\mu ]\rangle
\end{equation}
and the corresponding contact term ${\cal L}_\chi ^{J_1}$,
\begin{equation}
{\cal L}_\chi ^{J_1}=-\frac{f_\chi ^2}{2m^2}\langle [\chi _{-},u_\mu ][\chi
_{-},u_\mu ]\rangle\,,
\end{equation}
was added to ${\cal L}_{\chi ,T}^{(6)}$ by hand. In \cite{CEEKPP}, the same
was done with
\begin{equation}
J_1{}_\mu ^a=i\theta _V\varepsilon _{\mu \nu \alpha \beta }\left\langle
T^au^\nu u^\alpha u^\beta \right\rangle +h_V\varepsilon _{\mu \nu \alpha
\beta }\left\langle T^a\{u^\nu ,f_{+}^{\alpha \beta }\}\right\rangle
\end{equation}
and analogous source for the $1^{++}$ resonance nonet. In connection with this
{\em ad hoc} procedure\footnote{However, the addition of these contact terms to
${\cal L}_T$ can be dictated by the high-energy constraints as pointed out
in~\cite{CEEKPP}.}, a natural question arises whether there exist formulation
of the resonance Lagrangian, which produces {\em all} the terms in ${\cal
L}_\chi ^{(6)}$ known form both vector and antisymmetric tensor formulations
automatically. A closer look at the formula (\ref{L_RV}) suggests, that at
least the term ${\cal L}_\chi ^{J_1}$ together with all the terms produced by
the antisymmetric tensor field Lagrangian (\ref {start_L_T}) could be produced
starting with the first order Lagrangian ${\cal L}_{RV}$ (derived from ${\cal
L}_T$ by integrating in the vector field $V$)
\begin{eqnarray}
{\cal L}_{RV} &=&\frac 12m^2(V\cdot V)+m\left( J_1\cdot V\right) +\frac
12\left( J_1\cdot J_1\right)  \nonumber \\
&&+\frac 14m^2\left( R:R\right) +\left( J_2:R\right) +\left( R:J_4:R\right)
\nonumber \\
&&+\left( R:J_5\cdot D:R\right) -m\left( W\cdot V\right)\,.
\end{eqnarray}
provided we discard the contact term $\frac 12\left( J_1\cdot J_1\right) $.
The role of this term, which has to be present in ${\cal L}_{RV}$, is just
to cancel analogous term which could appear after the $R$ and $V$ fields
have been integrated out, because the tensor formalism does not produce such
a contact term and ${\cal L}_{RV}$ was constructed in order to be {\em %
completely} equivalent to ${\cal L}_T$.

\subsection{Basic properties}

Therefore we are encouraged to start with the following more general first
order Lagrangian
\begin{eqnarray}
{\cal L}_{VT} &=&\frac 14m^2(R:R)+\frac 12m^2\left( V\cdot V\right) -\frac
12m\left( R:\widehat{V}\right)  \nonumber \\
&&+\left( J_1\cdot V\right) +\frac 12\left( V\cdot K\cdot V\right) +\left(
V\cdot J_3:R\right)  \nonumber \\
&&+(R:J_4:R)+(R:J_5\cdot D:R)+\left( J_2:R\right)  \nonumber \\
&&+(R:J_6::RR)\,,  \label{L_VT}
\end{eqnarray}
where $V$ is vector field, $\widehat{V}=\widehat{DV}$ and $R$ is antisymmetric
tensor field. Note that in the notation we use, the term $-\frac 12m\left(
R:\widehat{V}\right) $ is equivalent (up to total derivative) to the term
$m(V\cdot W)$ with $W=D\cdot R$. The equations of motion indicate the chiral
counting $R=O(p^2)$ and $V=O(p^3)$, we have therefore
\begin{equation}
{\cal L}_{VT}={\cal L}_{VT}^{(4)}+{\cal L}_{VT}^{(6)}+{\cal L}_{VT}^{(8)}\,,
\end{equation}
where
\begin{eqnarray*}
{\cal L}_{VT}^{(4)} &=&\frac 14m^2(R:R)+\left( J_2^{(2)}:R\right)\,, \\
{\cal L}_{VT}^{(6)} &=&\frac 12m^2\left( V\cdot V\right) -\frac 12m\left( R:%
\widehat{V}\right) +\left( J_1\cdot V\right) +\left( J_2^{(4)}:R\right) \\
&&+\left( V\cdot J_3:R\right) +(R:J_4:R)+(R:J_5\cdot D:R)+(R:J_6::RR)\,, \\
{\cal L}_{VT}^{(8)} &=&\frac 12\left( V\cdot K\cdot V\right)\,.
\end{eqnarray*}
Solutions of the equations of motion to the lowest order read
\begin{eqnarray*}
R^{(2)} &=&-\frac 2{m^2}J_2^{(2)}\,, \\
V^{(3)} &=&-\frac 1{m^2}\left( J_1-\frac 2{m^2}J_3:J_2^{(2)}-\frac 2mD\cdot
J_2^{(2)}\right)\,.
\end{eqnarray*}
Inserting this to the original Lagrangian we obtain the effective chiral
Lagrangian up to the order $O(p^6)$ in the form
\begin{equation}\label{LchVT}
{\cal L}_{\chi ,VT}={\cal L}_{\chi ,VT}^{(4)}+{\cal L}_{\chi ,VT}^{(6)}\,,
\end{equation}
where
\begin{eqnarray*}
{\cal L}_{\chi ,VT}^{(4)} &=&-\frac 1{m^2}\left( J_2^{(2)}:J_2^{(2)}\right)\,,
\\
{\cal L}_{\chi ,VT}^{(6)} &=&-\frac 1{2m^2}\left( J_1\cdot J_1\right) -\frac
2{m^2}\left( J_2^{(2)}:J_2^{(4)}\right) -\frac
8{m^6}(J_2^{(2)}:J_6::J_2^{(2)}J_2^{(2)}) \\
&&+\frac 2{m^3}\left( D\cdot J_2^{(2)}\cdot J_1\right) +\frac 2{m^4}\left(
D\cdot J_2^{(2)}\cdot J_2^{(2)}\cdot \overleftarrow{D}\right) \\
&&-\frac 4{m^5}(D\cdot J_2^{(2)}\cdot J_3:J_2^{(2)})+\frac
4{m^4}(J_2^{(2)}:J_4:J_2^{(2)})+\frac 4{m^4}(J_2^{(2)}:J_5\cdot D:J_2^{(2)})
\\
&&-\frac 2{m^6}(J_2^{(2)}:J_3\cdot J_3:J_2^{(2)})+\frac 2{m^4}(J_1\cdot
J_3:J_2^{(2)})\,.
\end{eqnarray*}
Note that we have reproduced with the Lagrangian ${\cal L}_{VT}$ all the terms,
which yield the vector and tensor representation, and two new terms in the last
line. As an example of (\ref{LchVT}) we have calculated in Appendix \ref{A3}
the resonance saturation of some of the $O(p^6)$ LECs of the canonical basis
\cite{BCE} by means of ${\cal L}_{VT}$ with sources listed in Appendix
\ref{apJVT}. This can be compared with tensor formalism given in
\cite{moussallam}.

Note also that ${\cal L}_{\chi ,VT}^{(6)}$ can be rewritten in the more
condensed form
\begin{equation}
{\cal L}_{\chi ,VT}^{(6)}=-\frac 1{2m^2}\left( \widetilde{J}_1\cdot
\widetilde{J}_1\right) -\frac 2{m^2}\left( J_2^{(2)}:\widetilde{J}%
_2^{(4)}\right) +\frac 4{m^4}(J_2^{(2)}:J_5\cdot D:J_2^{(2)})\,,
\end{equation}
where
\begin{eqnarray*}
\widetilde{J}_1 &=&J_1-\frac2m J_2^{(2)}:J_3-\frac2m D\cdot J_2^{(2)}\,, \\
\widetilde{J}_2^{(4)} &=&J_2^{(4)}-\frac2{m^2} J_4:J_2^{(2)}+\frac4{m^4}
J_6::J_2^{(2)}J_2^{(2)}\,.
\end{eqnarray*}
This indicates some redundancy of the Lagrangian (\ref{L_VT}), as far as the
$O(p^6)$ effective chiral Lagrangian is concerned. This redundancy reflects
the possibility of field redefinitions according to the prescription
\begin{eqnarray}
R &\rightarrow& R-\frac 2{m^2}\delta J_4:R -\frac 2{m^2}V\cdot \delta ^{(1)}J_3 -\frac 2{m^2}\delta J_6::RR \,,\nonumber\\
V &\rightarrow &V-\frac 1{m^2}\delta ^{(2)}J_3:R \,,
\end{eqnarray}
which reproduces (up to the terms of the chiral order $O(p^8)$) the form of
the Lagrangian ${\cal L}_{VT}$ with the shifts of the sources
\begin{eqnarray}
J_1 &\rightarrow &J_1-\frac 2{m^2}\delta ^{(1)}J_3:J_2^{(2)}\,,  \nonumber \\
J_2^{(4)} &\rightarrow &J_2^{(4)}-\frac 2m\delta J_4:J_2^{(2)}-\frac
1{m^2}J_1\cdot \delta ^{(2)}J_3\,,  \nonumber \\
J_3 &\rightarrow &J_3-\delta ^{(1)}J_3-\delta ^{(2)}J_3\,,  \nonumber \\
J_4 &\rightarrow &J_4-\delta J_4-\frac 2{m^2}J_2^{(2)}:\delta J_6\,,  \nonumber
\\
J_5 &\rightarrow &J_5-\frac 12\widehat{g\delta ^{(2)}J}_3\,,  \nonumber \\
J_6 &\rightarrow &J_6-\delta J_6\,.  \label{J_transform}
\end{eqnarray}

\subsection{Correspondence with vector and antisymmetric tensor formalism}

Let us now establish the equivalence of this formalism with usual vector and
tensor representation. Writing
\begin{equation}
Z_{VT}[J_i,K]=\int {\cal D}R{\cal D}V\exp \left( i\int d^4x{\cal L}_{VT}\right)\,,
\end{equation}
we can integrate out either the vector field or the antisymmetric tensor
field and get
\begin{equation}
Z_{VT}[J_i,K]=\int {\cal D}R\exp \left( i\int d^4x{\cal L}_T^{\eff}\right)
=\int {\cal D}V\exp \left( i\int d^4x{\cal L}_V^{\eff}\right)\,.
\end{equation}
Using the formulae given in the preceding sections we end up with
\begin{eqnarray}
{\cal L}_T^{\eff} &=&\frac 14m^2\left( R:R\right) +\left( J_2:R\right)
+(R:J_4:R)+(R:J_5\cdot D:R) \nonumber\\
&&-\frac 1{2m^2}(J_1+R:J_3+mW)^2+(R:J_6::RR) \nonumber\\
&&-\frac 1{2m^2}\sum_{n=1}^\infty \left( -\frac 1{m^2}\right) ^n\left(
J_1+R:J_3+mW\right) \cdot K^n\cdot \left( J_1+R:J_3+mW\right)\,,
\end{eqnarray}
{\em i.e.}, up to $O(p^6)$
\begin{align}
{\cal L}_T^{\eff} =&-\frac 12\left( W\cdot W\right) +\frac 14m^2\left(
R:R\right)\notag \\
&+\left( J_1^{\eff}\cdot W\right) +\left( J_2^{\eff}:R\right) +\left( W\cdot
J_3^{\eff}:R\right) +\left( R:J_4^{\eff}:R\right) +\left( R:J_5^{\eff}\cdot
D:R\right) \notag\\
&+(R:J_6^{\eff}::RR)+{\cal L}_T^{\eff,contact}
\end{align}
with
$${\cal L}_T^{\eff,contact} =-\frac 1{2m^2}(J_1 \cdot J_1 )$$
and
\begin{alignat*}{3}
J_1^{\eff} &=-\frac 1mJ_1\,, &\qquad\qquad&
J_2^{\eff} &&=J_2-\frac 1{m^2}J_1\cdot J_3\,, \\
J_3^{\eff} &=-\frac 1mJ_3\,, &&
J_4^{\eff} &&=J_4-\frac 1{2m^2}J_3\cdot J_3\,, \\
J_5^{\eff} &=J_5\,, &&
J_6^{\eff} &&=J_6\,.
\end{alignat*}
As in the tensor case, the cubic term can be put off --
according to (\ref{J_transform}) we can  effectively absorbed the source $J_6$ in $J_4$.
We thus get
\begin{eqnarray}
{\cal L}_V^{\eff} &=&\frac 12m^2\left( V\cdot V\right) +\left( J_1\cdot
V\right) +\frac 12\left( V\cdot K\cdot V\right) \\
&&-\frac 1{m^2}(J_2-\frac 12m\widehat{V}+V\cdot J_3):(J_2-\frac 12m\widehat{V}+J_3\cdot V)
-\frac 1{m^2}\sum_{n=1}^\infty \left( -\frac 4{m^2}\right)^n\notag\\
&&\times(J_2-\frac 12m\widehat{V}+V\cdot J_3):(J_4-\frac 2{m^2}J_2^{(2)}:J_6+J_5\cdot
D)^n :(J_2-\frac 12m\widehat{V}+J_3\cdot V)\,,\notag
\end{eqnarray}
{\em i.e.} up to $O(p^6)$
\begin{equation}
{\cal L}_V^{\eff} =-\frac 14\left( \widehat{V}:\widehat{V}\right) +\frac
12m^2(V\cdot V) +(J_1^{\eff}\cdot V)+\left( J_2^{\eff}:\widehat{V}\right) +{\cal L}%
_V^{\eff,contact}
\end{equation}
with
\begin{eqnarray*}
{\cal L}_V^{\eff,contact} &=&-\frac 1{m^2}\left( J_2^{(2)}:J_2^{(2)}\right)
-\frac 2{m^2}\left( J_2^{(4)}:J_2^{(2)}\right) \\
&&+\frac 4{m^4}(J_2^{(2)}:J_4:J_2^{(2)})+\frac 4{m^4}(J_2^{(2)}:J_5\cdot
D:J_2^{(2)})-\frac 8{m^6}(J_2^{(2)}:J_6::J_2^{(2)}J_2^{(2)})
\end{eqnarray*}
and
\begin{alignat*}{3}
J_1^{\eff} &=J_1-\frac 2{m^2}J_2^{(2)}:J_3\,,&\qquad\qquad&
J_2^{\eff} &&=\frac 1mJ_2^{(2)}\,. \\
\end{alignat*}

\subsection{$VVP$ correlator}

Let us now return to the explicit example of the $VVP$ correlator. Within the
first order formalism we should take more general $J_i$ than in the vector and
tensor case. The reason is, that the integration by parts reducing the number
of independent terms re-distributing them between $J_1$ and $J_2$ and between
$J_3$ and $J_{4,5}$ (cf. (\ref{byparts})) is not possible here. More detailed
discussion as well as the explicit form of the sources is given in Appendix
\ref{apJVT}.

Using the Feynman rules described in the Appendix~\ref{appprop} and performing the
evaluation of the graphs depicted in Fig.~\ref{fig3} we get
\begin{align}
&\Pi_{VVP}^{VT}(p^2,q^2,r^2) =B_0\Bigl[ -\frac{N_C}{16\pi ^2r^2} +\frac{4\sqrt{2}h_Vp^2}{(p^2-m^2)r^2}\left( f_V-\frac{F_V}m\right)\notag\\
&\quad+2\left(F_V-f_V\frac{p^2}m\right) \left( F_V-f_V\frac{q^2}m\right)\frac{(d_1+8 d_2 -d_3)r^2+ 2 d_3 p^2}{(p^2-m^2)(q^2-m^2)r^2}\notag\\
&\quad+\frac{2\sqrt{2}}m\left( F_V-f_V\frac{p^2}m\right)
\frac{(c_1+c_2+8 c_3 -c_5)r^2+(-c_1+c_2+c_5-2c_6)p^2+(c_1-c_2+c_5)q^2}{(p^2-m^2)r^2}\notag\\
&\quad+\frac{2\sigma _Vm}{(p^2-m^2)(q^2-m^2)r^2}\left( f_V-\frac{F_V}m\right)
\left( F_V-f_V\frac{q^2}m\right) p^2 + (p \leftrightarrow q)\Bigr] .
\end{align}
Note that for $f_V=h_V=\sigma _V=0$ we reproduce $\Pi _{VVP}^T(p^2,q^2,r^2)$
(see eq. (\ref{VVP_T})) , while for $F_V=c_i=d_i=0$ we recover $\Pi
_{VVP}^V(p^2,q^2,r^2)$ (see eq. (\ref{VVP_V})).
\newsavebox{\propa}
\savebox{\propa}{\raisebox{-3pt}[0pt]{{\epsfig{file=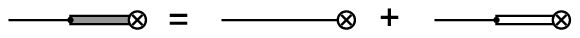}}}}
\newsavebox{\propb}
\savebox{\propb}{\raisebox{-6pt}[0pt]{{\epsfig{file=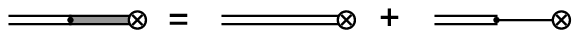}}}}
\begin{figure}[tb]
\begin{center}
\epsfig{file=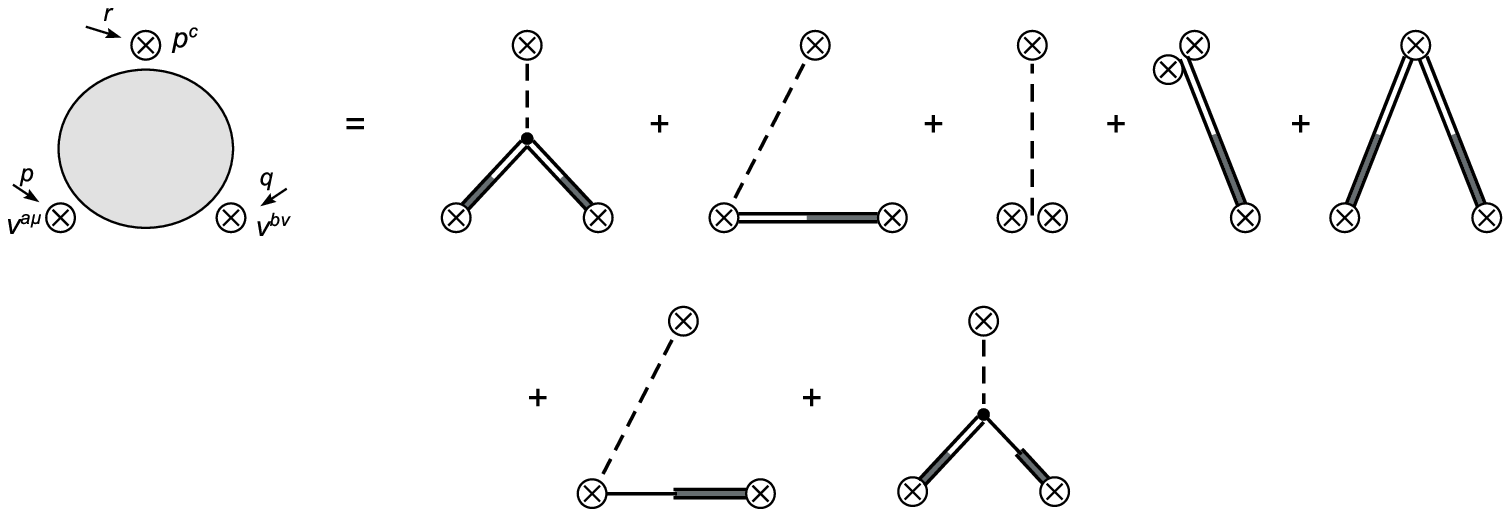}
\end{center}
\caption{Diagram representation of $VVP$ three point function in the first order
formalism. For limiting the number of graphs we have represented a pair of
resonance propagators by shaded double line as \usebox{\propa}\, and\, \usebox{\propb}\,.}
\label{fig3}
\end{figure}

The $VVP$ three point function $\Pi _{VVP}^{VT}(p^2,q^2,r^2)$ does not satisfy
the short distance constraint unless $f_V=0$ and the following relations for
the couplings hold\footnote{However, when appropriate $O(p^6)$ contact terms of
the form (\ref{VVP_CHPT}), (\ref{VVP_CHPT2}) are added to ${\cal L}_{VT}$ we
can fulfil the short distance constraints even for $f_V \neq 0$.}
\begin{eqnarray*}
c_1 &=&-4c_3\,, \\
c_2 &=&-4c_3+c_5\,, \\
c_6 &=&c_5-\frac{N_C}{64\sqrt{2}\pi ^2}\frac m{F_V}\,, \\
d_1 &=&-8d_2+\frac{\sigma _V}{2m}-\frac{N_C}{64\pi ^2}\frac{m^2}{F_V^2}+%
\frac{F_0^2}{4F_V^2}\,, \\
d_3 &=&\frac{\sigma _V}{2m}-\frac{N_C}{64\pi ^2}\frac{m^2}{F_V^2}+\frac{F_0^2%
}{8F_V^2}\,.
\end{eqnarray*}
Again the LMD approximation (\ref{LMD}) is reproduced in this case.

\section{Summary}\label{S6}

In this paper we have studied the formal relationship of the Proca field and
antisymmetric tensor field formalisms for describing massive spin-1
particles in the general context of chiral resonance Lagrangians. We have
concentrated on the various levels of equivalence of the corresponding
Lagrangians (\ref{L_V}), (\ref{L_T}) and tried to partially enlarge the
$O(p^4)$ analysis of ref. \cite{ecker2} to the order $O(p^6)$.

In this case, the situation is more involved, because at this order the
chiral resonance Lagrangian generally contain new types of bi(tri)linear
couplings of the resonance fields to the chiral building blocks. As a
result, contrary to the $O(p^4)$ case, the {\em complete equivalence}
(on the resonance level) of a local $O(p^6)$ Lagrangian within one formalism
cannot be generally achieved by means of a complementary finite chiral order
Lagrangian which is local and has finite number of interaction and contact
terms. Using the path integral method, previously used in this context in
\cite{bijnens} and \cite{tanabashi}, we have established the nonlocal
Lagrangians ${\cal L}_T^{\eff}$ and ${\cal L}_V^{\eff}$ (cf. (\ref
{L_T_eff_complet}), (\ref{L_Veff_Complete})) completely equivalent to ${\cal %
L}_V$ and ${\cal L}_T$ respectively.

On the other hand, provided we restrict ourselves to the $n-$point
correlators with $n\leq n_{\max }$, we can always reproduce them completely
with the complementary Lagrangian with finite number of terms and finite
chiral order $n_\chi \geq 6$. We have explicitly illustrated this using the
$VVP$ three point correlator as an explicit example.

Expanding the nonlocality of ${\cal L}_T^{\eff}$ and ${\cal L}_V^{\eff}$ and
keeping terms up to the chiral order $O(p^6)$ we have obtained local
Lagrangians ${\cal L}_T^{\eff,\,(\leq 6)}+{\cal L}_T^{\eff,\,(\leq 6)contact}$
and ${\cal L}_V^{\eff,\,(\leq 6)}+{\cal L}_V^{\eff,\,(\leq 6)contact}$ {\em %
strongly equivalent} to ${\cal L}_V$ and ${\cal L}_T$ respectively ({\em i.e.}
producing the same effective chiral Lagrangian up to $O(p^6)$).

There is also another aspect of the $O(p^6)$ chiral resonance Lagrangians we
have addressed. While at the order $O(p^4)$, the antisymmetric tensor
formalism is more complete in the sense that it produces all the possible
structures in the effective $O(p^4)$ chiral Lagrangian, at the order $O(p^6)$
neither the vector nor the antisymmetric tensor formalism has this property.
We have therefore suggested an alternative first order formalism, in the
framework of which the spin-1 particles are described in terms of a {\em pair%
} of vector and antisymmetric tensor fields. Within this formalism, all the
structures of the effective chiral Lagrangian which are known from the vector
and antisymmetric tensor representations are reproduced all at once with
certain additional terms. As we have shown in the special case of the $VVP$
correlator, provided we take the most general chiral building block to which
the vector and antisymmetric tensor fields can couple, we recover all the
contributions coming from the Lagrangians ${\cal L}_V$ and ${\cal L}_T$ at the
resonance level. This formalism also in some sense legitimizes the {\em ad hoc}
addition of the missing contact terms to the antisymmetric tensor formalism in
order to make the $O(p^6)$ effective chiral Lagrangian complete. Of course the
final decision justifying presence of such terms in ${\cal L}_T$ should by
dictated by asymptotic high energy constraints (cf. \cite{ecker2, CEEKPP}).

We have also studied the relationship of this first order representation to the
previous two. Again the completely equivalent Lagrangians ${\cal L}_T^{\eff}$
and ${\cal L}_V^{\eff}$ are generally nonlocal
and strongly equivalent local Lagrangians ${\cal L}_T^{\eff,\,(\leq 6)}+{\cal %
L}_T^{\eff,\,(\leq 6)contact}$ and ${\cal L}_V^{\eff,\,(\leq 6)}+{\cal L}%
_V^{\eff,\,(\leq 6)contact}$ can be constructed keeping terms up to the order
$O(p^6)$ only.

We have also briefly touched the issue of the short distance constraints
restricting the possible form of the chiral building blocks which appear in
the chiral resonance Lagrangian. For the $VVP$ correlator, which we have
studied in detail, it has been known \cite{RuizFemenia}, that the
antisymmetric tensor representation was compatible with the requirements
dictated by the leading order OPE, while the usual vector field
representation was not \cite{knecht}. However, the known correspondence of
the vector and antisymmetric representation can be applied here in order to
construct vector field chiral resonance Lagrangian, which for this special
correlator reproduces exactly the result of the antisymmetric tensor
formalism and therefore yield the result compatible with OPE. The price to
pay is the necessity to introduce terms up to the chiral order $O(p^{10})$.

As far as the first order formalism is concerned, because it is a synthesis
of the previous two, it can be therefore easily made compatible with the
short distance constraints for the $VVP$ correlator.

\section*{Acknowledgements}
We are grateful to J.~Ho\v{r}ej\v{s}\'{\i} and B.~Moussallam for careful
reading of the manuscript and helpful suggestions. The work was supported by
the Center for Particle Physics, project no. LC 527 of the Ministry of
Education of the Czech Republic.

\appendix

\section{Formal properties of the first order formalism} \label{appprop}

In this Appendix we summarize the properties of the first order formalism
introduced in Section~\ref{FO}. Let us consider the simplified case with
only one pair of (real) fields and the Lagrangian with two external sources
$J_{1,2}$
\begin{eqnarray}
{\cal L}_{VT} &=&\frac 14m^2R:R+\frac 12m^2V\cdot V-\frac 12mR:\widehat{V} \nonumber\\
&&+\left( J_1\cdot V\right) +\left( J_2:R\right)\,,
\end{eqnarray}
where $\widehat{V}=\widehat{\partial V}$. The equation of motion are then
\begin{eqnarray}
\frac 12m^2R-\frac 12m\widehat{V} &=&-J_2\,,  \nonumber \\
m^2V+mW &=&-J_1\,,  \label{free}
\end{eqnarray}
where $W=\partial \cdot R$. From the first equation it follows
\begin{equation}
\partial \cdot \widehat{V}-mW=\frac 2m\partial \cdot J_2
\end{equation}
and, when combined with the second one we get Proca equation for the $V$
field
\begin{equation}
\partial \cdot \widehat{V}+m^2V=\frac 2m\partial \cdot J_2-J_1\,.
\end{equation}
From the second eq. (\ref{free}) we get
\begin{equation}
\widehat{\partial W}+m\widehat{V}=-\frac 1m\widehat{\partial J_1}\,.
\end{equation}
This can be combined with the first eq. (\ref{free}) to obtain the standard
equation for the antisymmetric tensor field.
\begin{equation}
\widehat{\partial W}+m^2R=-\frac 1m\widehat{\partial J_1}-2J_2\,.
\end{equation}
We have therefore in the momentum representation (in our notation
$\widetilde{A}(p)=\int d^4x{\rm {e}}^{{\rm {i}}p\cdot x}A(x)$)
\begin{eqnarray}
\widetilde{V}(p) &=&-\Delta _F^V(p)\cdot \left( \widetilde{J}_1(p)+\frac{2%
{\rm i}}mp\cdot \widetilde{J}_2(p)\right)\,, \nonumber\\
\widetilde{R}(p) &=&-\Delta _F^R(p):\left( \widetilde{J}_2(p)-\frac{{\rm i}}{%
2m}\widehat{p\widetilde{J}_1(p)}\right)
\end{eqnarray}
and thus
\begin{equation}
\left(
\begin{array}{l}
\widetilde{V}(p) \\
\widetilde{R}(p)
\end{array}
\right) =-\left(
\begin{array}{ll}
\Delta _F^V(p) & -\frac{{\rm i}}m\widehat{\Delta _F^V(p)p} \\
\frac{{\rm i}}m\Delta _F^R(p)\cdot p & \Delta _F^R(p)
\end{array}
\right) \left(
\begin{array}{l}
\widetilde{J}_1(p) \\
\widetilde{J}_2(p)
\end{array}
\right)\,,
\end{equation}
where
\begin{eqnarray*}
{\rm i}\Delta _F^V(p)_{\mu \nu } &=&-\frac{{\rm i}}{p^2-m^2+{\rm i}0}\left(
g_{\mu \nu }-\frac{p_\mu p_\nu }{m^2}\right)\,, \\
{\rm i}\Delta _F^R(p)_{\mu \nu \,\,\rho \sigma } &=&-\frac{{\rm i}}{p^2-m^2+%
{\rm i}0}\frac 1{m^2}\left( (m^2-p^2)g_{\mu \rho }g_{\nu \sigma }+g_{\mu
\rho }p_\nu p_\sigma -g_{\mu \sigma }p_\nu p_\rho -(\mu \leftrightarrow \nu
)\right)\, .
\end{eqnarray*}
It is not difficult to prove, that
\begin{equation}
\Delta _F^{RV}(p)=-\frac{{\rm i}}m\widehat{\Delta _F^V(p)p}=-\frac{{\rm i}}%
m\Delta _F^R(p)\cdot p\,.
\end{equation}
The matrix of the momentum space propagators is therefore
\begin{equation}
\Delta _F(p)=\left(
\begin{array}{ll}
\Delta _F^V(p) & \Delta _F^{RV}(p) \\
\Delta _F^{RV}(-p)^T & \Delta _F^R(p)
\end{array}
\right)
\end{equation}
and the off-diagonal propagator is
\begin{equation}
i\Delta _F^{RV}(p)_{\sigma ,\mu \nu }=\frac{{\rm i}}{p^2-m^2+{\rm i}0}\frac{%
{\rm i}}m(g_{\sigma \mu }p_\nu -g_{\sigma \nu }p_\mu )\,.
\end{equation}
The two-point functions are then
\newsavebox{\propVV}
\savebox{\propVV}{\raisebox{2pt}[0pt]{{\epsfig{file=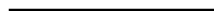}}}}
\newsavebox{\propRR}
\savebox{\propRR}{\raisebox{3pt}[0pt]{{\epsfig{file=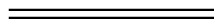}}}}
\newsavebox{\propVR}
\savebox{\propVR}{\raisebox{0pt}[0pt]{{\epsfig{file=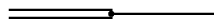}}}}
\newsavebox{\propRV}
\savebox{\propRV}{\raisebox{1pt}[0pt]{{\epsfig{file=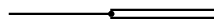}}}}
\begin{alignat}{3}
&\usebox{\propVV}\; &&= \langle T\widetilde{V}_\mu (p)V_\nu (0)\rangle &&={\rm i}\Delta _F^V(p)_{\mu
\nu }\,, \notag\\
&\usebox{\propRR} &&= \langle T\widetilde{R}_{\mu \nu }(p)R_{\rho \sigma }(0)\rangle &&={\rm i}
\Delta _F^R(p)_{\mu \nu \,\,\rho \sigma }\,, \notag\\
&\usebox{\propVR} &&= \langle T\widetilde{V}_\sigma (p)R_{\mu \nu }(0)\rangle &&={\rm i}\Delta
_F^{RV}(p)_{\sigma \,\,\mu \nu }\,, \notag\\
&\usebox{\propRV} && =  \langle T\widetilde{R}_{\mu \nu }(p)V_\sigma (0)\rangle &&={\rm i}\Delta
_F^{RV}(-p)_{\sigma \,\,\mu \nu }=-{\rm i}\Delta _F^{RV}(p)_{\sigma \,\,\mu
\nu }
\end{alignat}
and the wave functions to be attached to the on-shell external lines are
\begin{eqnarray}
\langle 0|V_\mu (0)|k\rangle &=&\varepsilon _\mu (k)\,, \nonumber\\
\langle 0|R_{\mu \nu }(0)|k\rangle &=&-\frac{{\rm i}}m\left( k_\mu
\varepsilon _\nu (k)-k_\nu \varepsilon _\mu (k)\right)\,.
\end{eqnarray}

\section{Structure of the $J_i$'s}
\label{apJ}

In this appendix we summarize explicit structure of sources $J_i$ needed in the
main text expressed by means of chiral building blocks. Our aim is not to give an
exhausting list of all $J_i$, instead we collect existing terms from the
literature. We preserve notation for couplings already developed there. In the
case of the first order formalism we combine both couplings from the vector and
tensor Lagrangians. Of course also in this case we do not attempt to give a
complete list.

In the following formulae, $T^a=\lambda ^a/\sqrt{2}$, where $\lambda ^a$ are
the Gell-Mann matrices and (in the next subsections) $\lambda^0 =
\sqrt{\frac23}\,\mathbf{1}$. The standard chiral building blocks are (for
details see {\em e.g.} \cite{BCE}).
\begin{eqnarray}
u &=&\exp \left( {\rm i}\phi ^aT^a/\sqrt{2}\right)\,,\nonumber  \\
u_\mu  &=&i[u^{\dagger}(\partial _\mu -ir_\mu )u-u(\partial _\mu -il_\mu )u^{\dagger}]\,,\nonumber \\
D_\mu  &=&\partial _\mu +\Gamma _\mu\,,\nonumber  \\
\Gamma _\mu  &=&\frac 12[u^{\dagger}(\partial _\mu -ir_\mu )u+u(\partial _\mu
-il_\mu )u^{\dagger}]\,,\nonumber \\
f_{\pm \mu \nu } &=&uf_{L\,\mu \nu }u^{\dagger}\pm u^{\dagger}f_{R\,\mu \nu }u\,,\nonumber \\
\chi _{\pm } &=&u^{\dagger}\chi u^{\dagger}\pm u\chi ^{\dagger}u\,,\nonumber \\
h_{\mu \nu } &=&D_\mu u_\nu +D_\nu u_\mu\,.
\end{eqnarray}

\subsection{Vector field formalism}
\label{apJV}

The $J_i$ corresponding to the vector field formulation (\ref{L_V}) we take
from \cite{Prades94} (see also \cite{knecht})

Parity even sector

\begin{itemize}
\item  $J_1$%
\[
\begin{tabular}{|c|c|c|}
\hline
 & Term $J_1{}_\mu ^a$ & coupling \\ \hline
1 & ${\rm i}\left\langle T^a[u^\nu ,f_{-\mu \nu }]\right\rangle $ & $\alpha
_V$ \\ \hline
2 & $\left\langle T^a[u_\mu ,\chi _{-}]\right\rangle $ & $\beta _V$ \\ \hline
\end{tabular}
\]

\item  $J_2$%
\[
\begin{tabular}{|c|c|c|}
\hline
& Term $J_2{}_{\mu \nu }^a$ & coupling \\ \hline
1 & $\left\langle T^af_{+\mu \nu }\right\rangle $ & $-\frac 1{2\sqrt{2}}f_V$
\\ \hline
2 & ${\rm i}\left\langle T^a[u_\mu ,u_\nu ]\right\rangle $ & $-\frac 1{2%
\sqrt{2}}g_V$ \\ \hline
\end{tabular}
\]

\item  $K$%
\[
\begin{tabular}{|c|c|c|}
\hline
& Term $K_{\mu \nu }^{ab}$ & coupling \\ \hline
1 & $g_{\mu \nu }\left\langle T^aT^bu_\alpha u^\alpha \right\rangle $ & $%
\delta _V^{(1)}$ \\ \hline
2 & $g_{\mu \nu }\left\langle T^au_\alpha T^bu^\alpha \right\rangle $ & $%
\delta _V^{(2)}$ \\ \hline
3 & $\left\langle T^aT^bu_\mu u_\nu \right\rangle $ & $\delta _V^{(3)}$ \\
\hline
4 & $\left\langle T^aT^bu_\nu u_\mu \right\rangle $ & $\delta _V^{(4)}$ \\
\hline
5 & $\left\langle T^au_\mu T^bu_\nu \right\rangle +\left\langle T^au_\nu
T^bu_\mu \right\rangle $ & $\delta _V^{(5)}$ \\ \hline
6 & $g_{\mu \nu }\left\langle T^aT^b\chi _{+}\right\rangle $ & $\kappa _V$
\\ \hline
7 & ${\rm i}\left\langle [T^a,T^b]f_{+\mu \nu }\right\rangle $ & $\phi _V$
\\ \hline
\end{tabular}
\]
\end{itemize}

Parity odd sector

\begin{itemize}
\item  $J_1$%
\[
\begin{tabular}{|c|c|c|}
\hline
& Term $J_1{}_\mu ^a$ & coupling \\ \hline
1 & ${\rm i}\varepsilon _{\mu \nu \alpha \beta }\left\langle T^au^\nu
u^\alpha u^\beta \right\rangle $ & $\theta _V$ \\ \hline
2 & $\varepsilon _{\mu \nu \alpha \beta }\left\langle T^a\{u^\nu
,f_{+}^{\alpha \beta }\}\right\rangle $ & $h_V$ \\ \hline
\end{tabular}
\]

\item  $J_3$%
\[
\begin{tabular}{|c|c|c|}
\hline
& Term $J_3{}_{\mu \nu }^{ab\alpha }$ & coupling \\ \hline
1 & $\varepsilon _{\,\,\,\beta \mu \nu }^\alpha \left\langle
\{T^a,T^b\}u^\beta \right\rangle $ & $\frac 12\sigma _V$ \\ \hline
\end{tabular}
\]
\end{itemize}

\subsection{Antisymmetric tensor field formalism}
\label{apJT}

Parity even sector to $O(p^4)$ (cf. \cite{ecker1} )

\begin{itemize}
\item  $J_2^{(2)}$%
\[
\begin{tabular}{|c|c|c|}
\hline
& Term $J_2{}_{\mu \nu }^a$ & coupling \\ \hline
1 & $\left\langle T^af_{+\mu \nu }\right\rangle $ & $\frac 1{2\sqrt{2}}F_V$
\\ \hline
2 & ${\rm i}\left\langle T^a[u_\mu ,u_\nu ]\right\rangle $ & $\frac 1{2\sqrt{%
2}}G_V$ \\ \hline
\end{tabular}
\]
\end{itemize}

Parity even sector $O(p^6)$ (here we give only the symmetry breaking terms
discussed in \cite{moussallam}; the complete list including also $J_6$ can be
found in \cite {CEEKPP})

\begin{itemize}
\item  $J_1$%
\[
\begin{tabular}{|c|c|c|}
\hline
& Term $J_1{}_\mu ^a$ & coupling \\ \hline
1 & $\left\langle T^a[\chi _{-},u_\mu ]\right\rangle $ & $-f_\chi / m$
\\ \hline
\end{tabular}
\]

\item  $J_2^{(4)}$%
\[
\begin{tabular}{|c|c|c|}
\hline
& Term $J_2{}_{\mu \nu }^a$ & coupling \\ \hline
1 & {\rm i}$\left\langle T^a\{[u_\mu ,u_\nu ],\chi _{+}\}\right\rangle $ & $%
\frac12 g_{V1}^m/m$ \\ \hline
2 & {\rm i}$\left\langle T^a(u_\mu \chi _{+}u_\nu -u_\nu \chi _{+}u_\mu
)\right\rangle $ & $\frac12 g_{V2}^m/m$ \\ \hline
3 & $\left\langle T^a\{f_{+\mu \nu },\chi _{+}\}\right\rangle $ &
$f_{V1}^m/m$ \\ \hline
4 & $\left\langle T^a[f_{-\mu \nu },\chi _{-}]\right\rangle $ &
$f_{V2}^m/m$ \\ \hline
\end{tabular}
\]

\item  $J_4$%
\[
\begin{tabular}{|c|c|c|}
\hline
& Term $J_4{}_{\mu \nu \,\,\alpha \beta }^{ab}$ & coupling \\
\hline
1 & $(g_{\mu \alpha }g_{\nu \beta }-g_{\nu \alpha }g_{\mu \beta
})\left\langle \chi _{+}\{T^a,T^b\}\right\rangle $ & $\frac 18e_V^m$ \\
\hline
\end{tabular}
\]
\end{itemize}

Parity odd sector to $O(p^6)$ (here we list the terms given in \cite
{RuizFemenia}, which are relevant for the calculation of the $VVP$ correlator;
cf. also \cite{pallante})

\begin{itemize}
\item  $J_2^{(4)}$%
\[
\begin{tabular}{|c|c|c|}
\hline & Term $J_2{}_{\mu \nu }^a$ & coupling \\ \hline 1 &
$\varepsilon _{\mu \kappa \rho \sigma }g_\nu ^\kappa \left\langle \{f_{+}^{\rho
\alpha },D_\alpha u^\sigma \}T^a\right\rangle $ & $c_1/m$
\\ \hline
2 & $\varepsilon _{\mu \kappa \rho \sigma }\left\langle \{f_{+}^{\rho \sigma
},D_\nu u^\kappa \}T^a\right\rangle $ & $c_2/m$ \\ \hline 3 &
$\varepsilon _{\mu \kappa \rho \sigma }g_\nu ^\kappa \left\langle
\{f_{+}^{\rho \sigma },\chi _{-}\}T^a\right\rangle $ & $c_3/m$ \\
\hline 4 & i$\varepsilon _{\mu \kappa \rho \sigma }g_\nu ^\kappa \left\langle
[f_{-}^{\rho \sigma },\chi _{+}]T^a\right\rangle $ & $c_4/m$ \\ \hline 5
& $\varepsilon _{\mu \kappa \rho \sigma }g_\nu ^\kappa \left\langle
D_\lambda \{f_{+}^{\rho \lambda },u^\sigma \}T^a\right\rangle $ & $-c_5/m$ \\
\hline 6 & $\varepsilon _{\mu \kappa \rho \sigma }\left\langle D_\nu
\{f_{+}^{\rho \sigma },u^\kappa \}T^a\right\rangle $ & $-c_6/m$ \\ \hline
7 & $\varepsilon _{\mu \kappa \rho \sigma }g_\nu ^\kappa \left\langle
D^\sigma \{f_{+}^{\rho \lambda },u_\lambda \}T^a\right\rangle $ & $-c_7/m$ \\ \hline
\end{tabular}
\]

\item  $J_4$%
\[
\begin{tabular}{|c|c|c|}
\hline
& Term $J_4{}_{\mu \nu \,\,\alpha \beta }^{ab}$ & coupling \\
\hline 1 & $\varepsilon _{\mu \nu \alpha \sigma }\left\langle D_\beta u^\sigma
\{T^a,T^b\}\right\rangle $ & $d_1$ \\ \hline 2 & $\varepsilon _{\mu \nu \alpha
\beta }\left\langle \chi _{-}\{T^a,T^b\}\right\rangle $ & $d_2$ \\ \hline
\end{tabular}
\]

\item  $J_5$%
\[
\begin{tabular}{|c|c|c|}
\hline
& Term $J_{5\,\mu \nu \,\,\rho \sigma }^{ab\,\alpha }$ &
coupling \\ \hline 1 & $\varepsilon _{\rho \sigma \mu \lambda }g_\nu ^\alpha
\left\langle \{T^a,T^b\}u^\lambda \right\rangle $ & $d_3$ \\ \hline 2 &
$\varepsilon _{\rho \sigma \mu \lambda }g^{\alpha \lambda }\left\langle
\{T^a,T^b\}u_\nu \right\rangle $ & $d_4$ \\ \hline
\end{tabular}
\]
\end{itemize}

\subsection{First order formalism}\label{apJVT}

For the first order formalism we can take $J_2^{(2)}$, $J_4$, $J_5$ and $J_6$
from the tensor formalism (see previous subsection).  As far as $K$ and $J_3$
are concerned we take them from the vector case, but in order to preserve the
dimension of the coupling $\sigma _V$, we have included one extra power of the
mass $m$ in the definition of $J_3$ (see below). The source $J_1$ incorporates
all the structure from the vector case supplemented with two additional terms
inferred from vector $J_2$ by means of integration by parts. Similarly for
parity even $J_2^{(4)}$, which we take from the tensor case while parity odd
$J_2^{(4)}$ stay unchanged. Though we preserve the notation from the vector and
tensor case, the phenomenological meaning of the couplings can be different.
Here we list only the differences:

Parity even $O(p^6)$

\begin{itemize}
\item  $J_1$%
\[
\begin{tabular}{|c|c|c|}
\hline
& Term $J_1{}_\mu ^a$ & coupling \\ \hline
1 & ${\rm i}\left\langle T^a[u^\nu ,f_{-\mu \nu }]\right\rangle $ & $\alpha
_V$ \\ \hline
2 & $\left\langle T^a[u_\mu ,\chi _{-}]\right\rangle $ & $\beta _V$ \\ \hline
3 & $\left\langle T^aD^\nu f_{+\nu \mu }\right\rangle $ & $\frac 1{\sqrt{2}%
}f_V$ \\ \hline
4 & ${\rm i}\left\langle T^aD^\nu [u_\nu ,u_\mu ]\right\rangle $ & $\frac 1{%
\sqrt{2}}g_V$ \\ \hline
\end{tabular}
\]
\item  $J_2^{(4)}$%
\[
\begin{tabular}{|c|c|c|}
\hline
& Term $J_2{}_{\mu \nu }^a$ & coupling \\ \hline
1 & {\rm i}$\left\langle T^a\{[u_\mu ,u_\nu ],\chi _{+}\}\right\rangle $ & $%
\frac12 g_{V1}^m/m$ \\ \hline
2 & {\rm i}$\left\langle T^a(u_\mu \chi _{+}u_\nu -u_\nu \chi _{+}u_\mu
)\right\rangle $ & $\frac12 g_{V2}^m/m$ \\ \hline
3 & $\left\langle T^a\{f_{+\mu \nu },\chi _{+}\}\right\rangle $ &
$f_{V1}^m/m$ \\ \hline
4 & $\left\langle T^a[f_{-\mu \nu },\chi _{-}]\right\rangle $ &
$f_{V2}^m/m$ \\ \hline
5 & $\left\langle T^a (D_\mu [\chi_-,\,u_\nu]-D_\nu [\chi_-,\,u_\mu]) \right\rangle $ &
$\frac12 f_\chi /m$ \\ \hline
\end{tabular}
\]

\end{itemize}

Parity odd $O(p^6)$

\begin{itemize}
\item  $J_1$%
\[
\begin{tabular}{|c|c|c|}
\hline
& Term $J_1{}_\mu ^a$ & coupling \\ \hline
1 & ${\rm i}\varepsilon _{\mu \nu \alpha \beta }\left\langle T^au^\nu
u^\alpha u^\beta \right\rangle $ & $\theta _V$ \\ \hline
2 & $\varepsilon _{\mu \nu \alpha \beta }\left\langle T^a\{u^\nu
,f_{+}^{\alpha \beta }\}\right\rangle $ & $h_V$ \\ \hline
\end{tabular}
\]

\item  $J_3$%
\[
\begin{tabular}{|c|c|c|}
\hline
& Term $J_3{}_{\mu \nu }^{ab\alpha }$ & coupling \\ \hline
1 & $\varepsilon _{\,\,\,\beta \mu \nu }^\alpha \left\langle
\{T^a,T^b\}u^\beta \right\rangle $ & $\frac 12m\sigma _V$ \\ \hline
\end{tabular}
\]
\end{itemize}

\section{Resonance contributions to the LECs}\label{A3}
This appendix summarizes saturation of the LECs of the
canonical basis given in~\cite{BCE} of the three flavour case ($n=3$ in the following table)
\begin{equation}
{\cal L}_6 = \sum_{i=1}^{94} C_i O_i\,,
\end{equation}
induced by the first order resonance Lagrangian ${\cal L}_{\chi ,VT}^{(6)}$ of
(\ref{LchVT}). The sources we have used are only those which are summarized in
the preceding appendix and are needed for constructions of chiral symmetry breaking terms.
Thus the saturation of LECs summarized in the next table is not complete and can be compared with
the similar results given in eqs.(62) of \cite{moussallam}.

\def\evm{ {e_V^m}}
\def\fvun{   {\, f^m_{V1}} }
\def\gvun{   {\, g^m_{V1}} }
\def\gvdeux{ {\, g^m_{V2}} }
\def\fvdeux{ {\, f^m_{V2}} }
\begin{longtable}[c]{|r|p{0.41\textwidth}||r|p{0.41\textwidth}|}
\hline
$ i $ & $C_i^R$ & $ i $ & $C_i^R$\\
\hline \hline
\endhead
\hline
\endfoot
\hline
1 & $\frac{G_V^2}{8 m^4}+\frac{g_V^2}{8 m^2}- \frac{G_V g_V}{4 m^3}$
&4& $\frac{G_V^2}{8 m^4}+\frac{g_V^2}{8 m^2}-\frac{G_V g_V}{4 m^3}$\\
5 & $- \frac{G_V g_{V2}^m}{\sqrt{2}m^3}$
&8& $\frac{\evm G_V^2}{2 m^4} - \frac{\sqrt{2} G_V  g_{V1}^m}{m^3}$\\
10& $-\frac{\evm G_V^2}{2 m^4}+ \frac{\sqrt{2} G_V g_{V1}^m}{m^3} +\frac{G_V g_{V2}^m}{\sqrt{2}m^3}$
&22&$\frac{G_V^2}{16 m^4} + \frac{G_V f_\chi}{2\sqrt 2 m^3}+\frac{g_V \beta_V}{2\sqrt2 m^2}+\frac{g_V^2}{16 m^2}-\frac{G_V \beta_V}{2\sqrt2 m^3}-\frac{G_V g_V}{8m^3}$\\
24& $\frac{G_V^2}{4n m^4}+\frac{g_V^2}{4 n m^2}-\frac{G_V g_V}{2 n m^3}$
&25&$-\frac{3 G_V^2}{8 m^4}-\frac{G_V f_\chi}{\sqrt{2}m^3}-\frac{g_V \beta_V}{\sqrt2 m^2}-\frac{3 g_V^2}{8 m^2}+\frac{G_V \beta_V}{\sqrt2 m^3}+\frac{3 G_V g_V}{4 m^3}$\\
26& $(1-\frac{2}{n^2})\frac{G_V^2}{4m^4}+\frac{G_V f_\chi}{\sqrt{2}m^3} + \frac{\beta_V^2}{m^2}+\frac{g_V \beta_V}{\sqrt2 m^2}+(1-\frac{2}{n^2})\frac{g_V^2}{4 m^2}-\frac{G_V \beta_V}{\sqrt2 m^3}-(1-\frac{2}{n^2})\frac{G_V g_V}{2m^3}$
&27&$-(\frac1n - \frac{2}{n^2})\frac{G_V^2}{4m^4}-(\frac1n - \frac{2}{n^2})\frac{g_V^2}{4 m^2}+(\frac1n-\frac{2}{n^2})\frac{G_V g_V}{2m^3}$\\
28& $\frac{G_V^2}{8n^2 m^4}+\frac{g_V^2}{8 n^2 m^2}-\frac{G_V g_V}{4 n^2 m^3}$
&29&$-(1+\frac{2}{n^2})\frac{G_V^2}{8 m^4}-\frac{G_V f_\chi}{\sqrt{2} m^3}-\frac{\beta_V^2}{m^2}-\frac{g_V \beta_V}{\sqrt2 m^2}-(1+\frac{2}{n^2})\frac{g_V^2}{8 m^2}+\frac{G_V \beta_V}{\sqrt2 m^3}+(1+\frac{2}{n^2})\frac{G_V g_V}{4m^3}$\\
30& $\frac{G_V^2}{4n^2 m^4}+\frac{g_V^2}{4 n^2 m^2}-\frac{G_V g_V}{2n^2 m^3}$
&40&$-\frac{G_V^2}{8 m^4}-\frac{g_V^2}{8 m^2}+\frac{G_V g_V}{4m^3}$\\
42& $-\frac{G_V^2}{8 m^4}-\frac{g_V^2}{8 m^2}+\frac{G_V g_V}{4 m^3}$
&44&$\frac{G_V^2}{4 m^4}+\frac{g_V^2}{4 m^2}-\frac{G_V g_V}{2m^3}$\\
48& $-\frac{G_V^2}{8 m^4}-\frac{g_V^2}{8 m^4}+\frac{G_V g_V}{4m^3}$
&50&$\frac{F_V G_V}{4 m^4}+ \frac{F_V f_\chi}{\sqrt2 m^3}+\frac{f_V \beta_V}{\sqrt2 m^2}+\frac{f_V g_V}{4 m^2}-\frac{F_V \beta_V}{\sqrt2 m^3}-\frac{F_V g_V}{4 m^3} - \frac{G_V f_V}{4 m^3}$\\
51& $-\frac{G_V^2}{4 m^4}+\frac{F_V G_V}{4 m^4} +\frac{F_V f_\chi}{\sqrt 2 m^3}+\frac{f_V \beta_V}{\sqrt2 m^2}+\frac{f_V g_V}{4 m^2}-\frac{g_V^2}{4 m^2}-\frac{F_V \beta_V}{\sqrt2 m^3}-\frac{F_V g_V}{4 m^3} - \frac{G_V f_V}{4 m^3}+\frac{G_V g_V}{2m^3}$
&52&$-\frac{F_V G_V}{4 m^4}-\frac{F_V f_\chi}{\sqrt2 m^3}-\frac{f_V \beta_V}{\sqrt2 m^2}-\frac{f_V g_V}{4 m^2}+\frac{F_V \beta_V}{\sqrt2 m^3}+\frac{F_V g_V}{4 m^3} + \frac{G_V f_V}{4 m^3}$\\
53& $-\frac{F_V G_V}{8 m^4}-\frac{3 F_V^2}{16 m^4} -\frac{F_V f_\chi}{2 \sqrt2 m^3}-\frac{f_V \beta_V}{2\sqrt2 m^2}- \frac{3 f_V^2}{16 m^2}-\frac{f_V g_V}{8 m^2}+\frac{F_V \beta_V}{2\sqrt2 m^3}+\frac{3 f_V F_V}{8 m^3}+\frac{F_V g_V}{8 m^3} + \frac{G_V f_V}{8 m^3}$
&55&$\frac{F_V G_V}{8 m^4} + \frac{3 F_V^2}{16 m^4} + \frac{F_V f_\chi}{2\sqrt 2 m^3}+\frac{f_V \beta_V}{2\sqrt2 m^2}+\frac{3 f_V^2}{16 m^2}+\frac{f_V g_V}{8 m^2}-\frac{F_V \beta_V}{2\sqrt2 m^3}-\frac{3 f_V F_V}{8m^3}-\frac{F_V g_V}{8 m^3} - \frac{G_V f_V}{8 m^3}$\\
56& $-\frac{F_V G_V}{4 m^4}+\frac{3 F_V^2}{8 m^4} - \frac{F_V f_\chi}{\sqrt2 m^3}-\frac{f_V \beta_V}{\sqrt2 m^2}+\frac{3 f_V^2}{8 m^2}-\frac{f_V g_V}{4 m^2}+\frac{F_V \beta_V}{\sqrt2 m^3}-\frac{3 f_V F_V}{4m^3}+\frac{F_V g_V}{4 m^3} + \frac{G_V f_V}{4 m^3}$
&57&$\frac{F_V G_V}{2 m^4}+\frac{F_V^2}{8 m^4} + \frac{\sqrt2 F_V f_\chi}{m^3}+\frac{2 f_V \beta_V}{\sqrt2 m^2}+\frac{f_V^2}{8 m^2}+\frac{f_V g_V}{2 m^2}-\frac{2 F_V \beta_V}{\sqrt2 m^3}-\frac{f_V F_V}{4 m^3}-\frac{F_V g_V}{2 m^3} - \frac{G_V f_V}{2 m^3}$\\
59& $-\frac{F_V G_V}{8 m^4} -\frac{F_V^2}{4 m^4} - \frac{F_V f_\chi}{2 \sqrt2 m^3}-\frac{f_V \beta_V}{2\sqrt2 m^2}-\frac{f_V^2}{4 m^2}-\frac{f_V g_V}{8 m^2}+\frac{F_V \beta_V}{2\sqrt2 m^3}+\frac{f_V F_V}{2 m^3}+\frac{F_V g_V}{8 m^3} + \frac{G_V f_V}{8 m^3}$
&61&$\frac{\evm F_V^2}{4 m^4} - \frac{\sqrt2 F_V \fvun}{m^3}$\\
63& $-\frac{\sqrt2 \fvun G_V}{m^3} + \frac{\evm F_V G_V}{2 m^4} - \frac{F_V \gvun}{\sqrt2 m^3}$
&65&$-\frac{F_V \gvdeux}{\sqrt2 m^3}$\\
66& $\frac{G_V^2}{8 m^4}+\frac{g_V \alpha_V}{2\sqrt{2}m^2}+\frac{g_V^2}{8 m^2}-\frac{G_V \alpha_V}{2\sqrt2 m^3}-\frac{G_V g_V}{4m^3}$
&69&$-\frac{G_V^2}{8 m^4}-\frac{g_V \alpha_V}{2\sqrt{2}m^2}-\frac{g_V^2}{8 m^2}+\frac{G_V \alpha_V}{2\sqrt2 m^3}+\frac{G_V g_V}{4m^3}$\\
70& $-\frac{G_V^2}{8 m^4}-\frac{F_V G_V}{8 m^4} +\frac{F_V^2}{8 m^4} - \frac{F_V f_\chi}{2 \sqrt2 m^3} -\frac{f_V \beta_V}{2\sqrt2 m^2}+\frac{f_V^2}{8 m^2}-\frac{f_V g_V}{8 m^2}-\frac{g_V^2}{8 m^2}+\frac{F_V \beta_V}{2\sqrt2 m^3}-\frac{f_V F_V}{4 m^3}+\frac{F_V g_V}{8 m^3} + \frac{G_V f_V}{8 m^3}+\frac{G_V g_V}{4m^3}$
&72&$\frac{F_V G_V}{8 m^4} -\frac{F_V^2}{8 m^4} + \frac{F_V f_\chi}{2\sqrt 2 m^3}+\frac{f_V \beta_V}{2\sqrt2 m^2}-\frac{f_V^2}{8 m^2}+\frac{f_V g_V}{8 m^2}-\frac{F_V \beta_V}{2\sqrt2 m^3}+\frac{f_V F_V}{4 m^3}-\frac{F_V g_V}{8 m^3} - \frac{G_V f_V}{8 m^3}$\\
73& $\frac{F_V G_V}{4 m^4}-\frac{F_V^2}{8 m^4} + \frac{F_V f_\chi}{\sqrt2 m^3}+\frac{f_V \beta_V}{\sqrt2 m^2}-\frac{f_V^2}{8 m^2}+\frac{f_V g_V}{4 m^2}-\frac{F_V \beta_V}{\sqrt2 m^3}+\frac{f_V F_V}{4m^3}-\frac{F_V g_V}{4 m^3} - \frac{G_V f_V}{4 m^3}$
&74&$-\frac{G_V^2}{4 m^4}-\frac{\alpha_V^2}{m^2}-\frac{g_V \alpha_V}{\sqrt{2}m^2}-\frac{g_V^2}{4 m^2}+\frac{G_V \alpha_V}{\sqrt2 m^3}+\frac{G_V g_V}{2m^3}$\\
76& $-\frac{F_V G_V}{8 m^4}+\frac{F_V^2}{16 m^4} - \frac{F_V f_\chi}{2\sqrt2 m^3} + \frac{\alpha_V^2}{2 m^2}+\frac{g_V \alpha_V}{2\sqrt{2}m^2}-\frac{f_V \beta_V}{2\sqrt2 m^2}+\frac{f_V^2}{16 m^2}-\frac{f_V g_V}{8 m^2}+\frac{F_V \beta_V}{2\sqrt2 m^3}-\frac{f_V F_V}{8m^3}+\frac{F_V g_V}{8 m^3} + \frac{G_V f_V}{8 m^3}-\frac{G_V \alpha_V}{2\sqrt2 m^3}$
&78&$ \frac{F_V G_V}{8 m^4} +\frac{F_V^2}{4 m^4} + \frac{F_V f_\chi}{2\sqrt2 m^3}+\frac{f_V \beta_V}{2\sqrt2 m^2}+\frac{f_V^2}{4 m^2}+\frac{f_V g_V}{8 m^2}-\frac{F_V \beta_V}{2\sqrt2 m^3}-\frac{f_V F_V}{2m^3}-\frac{F_V g_V}{8 m^3} - \frac{G_V f_V}{8 m^3}$\\
79& $-\frac{F_V G_V}{8 m^4} + \frac{F_V^2}{8 m^4} - \frac{F_V f_\chi}{2\sqrt2 m^3}-\frac{f_V \beta_V}{2\sqrt2 m^2}+\frac{f_V^2}{8 m^2}-\frac{f_V g_V}{8 m^2}+\frac{F_V \beta_V}{2\sqrt2 m^3}-\frac{f_V F_V}{4m^3}+\frac{F_V g_V}{8 m^3} + \frac{G_V f_V}{8 m^3}$
&82&$-\frac{F_V G_V}{16 m^4} -\frac{F_V^2}{16 m^4} - \frac{F_V f_\chi}{4\sqrt2 m^3} - \frac{\fvdeux F_V}{\sqrt2 m^3}-\frac{f_V \beta_V}{4\sqrt2 m^2}-\frac{f_V^2}{16 m^2}-\frac{f_V g_V}{16 m^2}+\frac{F_V \beta_V}{4\sqrt2 m^3}+\frac{f_V F_V}{8m^3}+\frac{F_V g_V}{16 m^3} + \frac{G_V f_V}{16 m^3}$\\
83& $ \frac{3 G_V^2}{16 m^4} + \frac{G_V f_\chi}{2\sqrt2 m^3}-\frac{\sqrt2 \fvdeux G_V}{m^3}+\frac{g_V \alpha_V}{2\sqrt{2}m^2}+ \frac{\alpha_V\beta_V}{m^2}+\frac{g_V \beta_V}{2\sqrt2 m^2}+\frac{3 g_V^2}{16 m^2}-\frac{G_V \alpha_V}{2\sqrt2 m^3}-\frac{G_V \beta_V}{2\sqrt2 m^3}-\frac{3 G_V g_V}{8m^3}$
&87&$ \frac{F_V^2}{8 m^4}+\frac{f_V^2}{8 m^2}-\frac{f_V F_V}{4m^3}$\\
88& $-\frac{F_V G_V}{4 m^4} -\frac{F_V f_\chi}{\sqrt2 m^3}-\frac{f_V \beta_V}{\sqrt2 m^2}-\frac{f_V g_V}{4 m^2}+\frac{F_V \beta_V}{\sqrt2 m^3}+\frac{F_V g_V}{4 m^3} + \frac{G_V f_V}{4 m^3}$
&89&$\frac{F_V^2}{2 m^4} + \frac{F_V G_V}{4 m^4} + \frac{\alpha_V f_V}{\sqrt{2} m^2}+ \frac{f_V^2}{2 m^2}+\frac{f_V g_V}{4 m^2}-\frac{F_V \alpha_V}{\sqrt2 m^3}-\frac{f_V F_V}{m^3}-\frac{F_V g_V}{4 m^3} - \frac{G_V f_V}{4 m^3}$\\
90& $-\frac{F_V f_\chi}{\sqrt2 m^3}-\frac{f_V \beta_V}{\sqrt2 m^2}+\frac{F_V \beta_V}{\sqrt2 m^3}$
&92&$\frac{F_V^2}{m^4}+\frac{f_V^2}{m^2}-\frac{2 f_V F_V}{m^3}$\\
93& $-\frac{F_V^2}{4 m^4}-\frac{f_V^2}{4 m^2}+\frac{f_V F_V}{2 m^3}$&&
\end{longtable}

\end{document}